\documentclass[11pt]{article}
\usepackage{amsfonts}
\usepackage{tikz}
\usepackage{graphicx}
\usepackage{bbold}
\usepackage{multirow}
\usepackage[utf8]{inputenc}
\usepackage[T1]{fontenc}
\usepackage{url}
\usepackage{amsmath}

\usetikzlibrary{automata,positioning}
\newtheorem{theorem}{Theorem}

\newtheorem{lemma}{Lemma}

\newtheorem{proposition}{Proposition}
\newtheorem{remark}{Remark}

\newenvironment{proof}[1][Proof]{\noindent \textbf{#1.} }{\  \rule{0.5em}{0.5em}}
\input{tcilatex}

\makeatletter
\renewcommand\@biblabel[1]{}
\renewenvironment{thebibliography}[1]
     {\section*{\refname}%
      \@mkboth{\MakeUppercase\refname}{\MakeUppercase\refname}%
      \list{}%
           {\leftmargin0pt
            \@openbib@code
            \usecounter{enumiv}}%
      \sloppy
      \clubpenalty4000
      \@clubpenalty \clubpenalty
      \widowpenalty4000%
      \sfcode`\.\@m}
     {\def\@noitemerr
       {\@latex@warning{Empty `thebibliography' environment}}%
      \endlist}
\makeatother

\rm 
\allowdisplaybreaks

\begin{document}

\title{Statistical inference for a mixture of Markov jump processes}
\author{Halina Frydman \thanks{%
Department of Technology, Operations and Statistics, Stern School of
Business, New York University, New York, NY 10012, USA, email:
hfrydman@stern.nyu.edu} \, Budhi Surya \thanks{%
School of Mathematics and Statistics, Victoria University of Wellington,
Gate 6, Kelburn PDE, Wellington 6140, New Zealand, email:
budhi.surya@vuw.ac.nz} \thanks{%
The author acknowledges financial support through PBRF research grant No.
220859.} }
\date{10 April 2022}
\maketitle

\begin{abstract}
We estimate a general mixture of Markov jump processes. The key novel
feature of the proposed mixture is that the transition intensity matrices of
the Markov processes comprising the mixture are entirely unconstrained. The
Markov processes are mixed with distributions that depend on the initial
state of the mixture process. The new mixture is estimated from its
continuously observed realizations using the EM algorithm, which provides
the maximum likelihood (ML) estimates of the mixture's parameters. We derive
the asymptotic properties of the ML estimators. To obtain estimated standard
errors of the ML estimates of the mixture's parameters, an explicit form of
the observed Fisher information matrix is derived. In its new form, the
information matrix simplifies the conditional expectation of outer product
of the complete-data score function in the Louis (1982) general matrix
formula for the observed Fisher information matrix. Simulation study
verifies the estimates' accuracy and confirms the consistency and asymptotic
normality of the estimators. The developed methods are applied to a medical
dataset, for which the likelihood ratio test rejects the constrained mixture
in favor of the proposed unconstrained one. This application exemplifies the
usefulness of a new unconstrained mixture for identification and
characterization of homogeneous subpopulations in a heterogeneous population.


\textbf{keywords}: mixture of Markov jump processes, EM algorithm, Fisher
information matrix, asymptotic distribution, heterogeneous population
\end{abstract}

\pagestyle{myheadings} 
\markboth{Frydman, Surya}{Frydman, Surya: Statistical
inference for a mixture of Markov jump processes}

\section{Introduction}

This paper proposes a general mixture of Markov jump processes. This new
mixture is an extension of the model in Frydman (2005), which in turn
extends the seminal mover-stayer model presented in Blumen, \textit{et al}.
(1955). The key novel feature of the proposed model is that the transition
intensity matrices of the Markov processes comprising the mixture are
entirely unconstrained: each homogeneous subpopulation evolves according to
a Markov process with its own intensity matrix. The mixture's regime
membership distribution is assumed to depend only on the initial state of
the process, and may differ between initial states. Because constraining the
transition intensities may obscure the identification of clusters, the
proposed unconstrained mixture is particularly suitable for identification
and characterization of homogeneous subpopulations in a heterogeneous
population. This is illustrated with a medical application in which the
likelihood ratio test rejects the constrained mixture from Frydman (2005) in
favor of the proposed general mixture.

We obtain the maximum likelihood (ML) estimates of a general mixture
(g-mixture) from the data consisting of a set of its continuously observed
realizations using the EM algorithm of Dempster, \textit{et al}. (1977).
Using novel methods, we derive the asymptotic properties of the ML
estimators thereby extending the classic results from Albert (1962) who
considered ML estimation of a single Markov jump process. For a finite
sample of realizations, we derive an explicit form of the
observed Fisher information matrix to obtain an estimate of the covariance
matrix of the MLEs. In its new form, the information matrix simplifies the
conditional expectation of outer product of the complete-data score function
in Louis's (1982) general matrix formula for the observed Fisher information
matrix. We show through the simulation study that the estimation is
accurate and confirms the asymptotic properties of the estimators. We note
that Frydman (2005) provided the MLEs of the parameters of a constrained
mixture defined below, but hasn't considered their asymptotic or finite
sample properties\textit{. }The methods developed here are applied to the
ventICU dataset from Cook and Lawless (2018).

The g-mixture is different from the mixture of Markov processes recently
considered by Jiang and Cook (2019). There, the regime probabilities depend
on covariates through the multinomial logistic regression, while the Markov
processes in the mixture are assumed to have the same intensity matrices. In
the g-mixture, the regime probabilities depend only on an initial state,
while the Markov processes have their own intensity matrices. We observe
g-mixture continuously, whereas Jiang and Cook (2019) observe their mixture
intermittently.

To define the proposed model, let $X=\{X_{m},1\leq m\leq M\}$ be the mixture
of $M$ right-continuous Markov jump processes with the intensity matrices $%
Q_{m}^{\prime }s,$ and transition matrices $P_{m}(t)=\exp (Q_{m}t)$ defined
on the finite state space $S=E\cup \Delta ,$ where $E$ is a set of non
absorbing states and $\Delta $ a set of absorbing states. There is a
separate mixing distribution for each initial state $i\in E$ of $X,$ 
\begin{equation*}
\phi _{i,m}\equiv \mathbb{P}(X=X_{m}|X_{0}=i),\;\;1\leq m\leq M,
\end{equation*}%
where $\sum_{m=1}^{M}\phi _{i,m}=1.$ Let $D_{m}=$diag$(\phi _{1,m},...,\phi
_{w,m}),$ where $w$ is the cardinality of $E.$ Then the transition matrix of
a mixture process $X$ is given by 
\begin{equation*}
P(t)=\sum_{m=1}^{M}D_{m}P_{m}(t),\;\;t\geq 0.
\end{equation*}%
In the absence of absorbing states, Frydman (2005) specified the following
structure for the intensity matrices of Markov processes comprising the
mixture%
\begin{equation*}
Q_{m}=\Gamma _{m}Q\text{ }\left( 1\leq m\leq M\right) ,
\end{equation*}%
where $Q$ is an intensity matrix, $\Gamma _{m}=$diag$(\gamma
_{1,m},...\gamma _{w,m}),$ with $\gamma _{i,m}\geq 0,$ for $1\leq m\leq M-1,$
and $\Gamma _{M}=I,$ an identity matrix$.$ Depending on whether $\gamma
_{i,m}=0,$ $0<\gamma _{i,m}<1,$ $\gamma _{i,m}\geq 1,$ the realizations
generated by $Q_{m}$ do not move out of state $i$, or move out of state $i$
at a lower or higher rate than those generated by $Q$, or at an identical
rate. This specification constrains the transition matrices of Markov chains
embedded into Markov processes in the mixture to be all the same. This
mixture is restrictive in situations in which the population is
heterogeneous not only with respect to exit rates from states, but also with
respect to the direction of movement. As illustrated in the medical
application in this paper, it is particularly restrictive when the
components of the mixture are absorbing Markov processes as in this case the
absorption probabilities would be the same for all Markov processes in the
mixture. Nevertheless, this mixture has been successfully applied to
modeling bond-ratings migration by Frydman and Schuermann (2008), and to
clustering of categorical time series by Pamminger and Fruhwirth-Schnatter
(2010). Its distributional properties were studied in Surya (2018).

By setting $M=2,$ and $\gamma _{1,1}=...=\gamma _{w,1}=0$ in $\Gamma _{1}$,
the transition matrix of ${X}$ reduces to $D_{1}I+(I-D_{1})P_{2}(t)$, a
transition matrix of a continuous-time mover-stayer (MS) model. The MS model
assumes a simple form of population heterogeneity: there are stayers who
never leave their initial states, with $P_{1}(t)=I$ as their transition
matrix, and movers who evolve among the states according to transition
matrix $P_{2}(t)=\exp (tQ)$. The MS model was the first mixture of Markov
chains considered in the literature, and its use in Blumen, \textit{et al}.
(1955) to study labor mobility was the first application of stochastic
processes in the social sciences. Frydman (1984) obtained the ML estimators
of the discrete-time MS model's parameters by direct maximization of the
observed likelihood function, and Fuchs and Greenhouse (1988) did so by
using the expectation-maximization (EM) algorithm. In both cases, estimation
used the data on independent realizations of the MS model.

Despite capturing a very simple form of population heterogeneity, the MS
mixture's discrete and continuous-time versions have been widely applied in
diverse fields, including medicine (Tabar, \textit{et al}., 1996), labor
economics (Fougere and Kamionka, 2003), large data (Cipollini, \textit{et al}%
., 2012), farming (Saint-Cyr and Piet, 2017), and credit risk (Frydman and
Kadam, 2004, and Ferreti, \textit{et al.}, 2019). In a continuous-time
framework, Yi, \textit{et al}. (2017) estimated an MS model from panel data
in the presence of state misclassification. Cook, et al. (2002) developed a
generalized MS model, which allows for subject specific absorbing states,
and Shen and Cook (2014) considered a dynamic MS model for recurrent events
that can be resolved. In a discrete-time framework, Frydman and Matuszyk
(2018, 2019) developed an estimation for a discrete-time MS model with
covariate effects on stayers' probability and movers' transitions.

The paper is organized as follows. Section 2 sets the notation and derives
the observed and complete likelihood functions. Section 3 presents the EM
algorithm, the derivation of the asymptotic properties of the MLEs and also
provides the lower bound for the asymptotic variance of the MLEs. The finite
sample covariance matrix of the MLEs is derived in Section 4. Section 5 is
devoted to the simulation study and Section 6 applies the developed methods
to ventICU data. Section 7 concludes the paper.

\section{The observed and complete likelihood functions}

\subsection{The observed likelihood function}

We consider the general continuous-time mixture $X$ with $M$ components
defined in the Introduction. Let $X^{k}=\{X_{t}^{k},0\leq t\leq T\}$ denote
the k'th realization of $X$ on $[0,T]$ where $T$ is the end-of-study time,
which can be either fixed, or the absorption time of $X$. Denote by $X^{k}-$
that realization without an initial state, that is, $X^{k}=(X^{k}-)\cup
X_{0}.$ And, let $R_{k}$ denote the regime label of the k'th realization. To
write the observed likelihood of $X^{k},$ for $1\leq k\leq K$ and $1\leq
m\leq M,$ we define the following quantities associated with $X^{k}$: 
\begin{eqnarray}
\Phi _{k,m} &=&I(R_{k}=m)  \notag \\
B_{i}^{k} &=&I(X_{0}^{k}=i),i\in E,  \notag \\
B_{i} &=&\sum_{k=1}^{K}B_{i}^{k}=\#\text{ of realizations with initial state 
}i,i\in E  \label{a0} \\
N_{ij}^{k} &=&\#\text{ of times }X^{k}\text{ makes an }i\rightarrow j\text{
transition, }i\neq j,i\in E  \notag \\
T_{i}^{k} &=&\int_{0}^{T}I(X_{u}^{k}=i)du=\text{ total time }X^{k}\text{
spends in state }i\in E,  \notag
\end{eqnarray}%
where $\Phi _{k,m}$ is equal to 1 if the k'th realization evolves according
to the m'th Markov process and equal to zero, otherwise. We note that $\Phi
_{k,m}$ is unknown, but other quantities in (\ref{a0}) are known. For $1\leq
m\leq M,$ let $\theta _{m}\equiv (q_{ij,m}$,$i\neq j,$ $\phi _{i,m},i\in
E,j\in S)$ and $\theta =(\theta _{m},1\leq m\leq M)$ be the mixture's
parameters assumed to live on the compact set $\Theta$ of any positive value
of $\theta$. We denote by $\mathbb{P}_{\theta}$ the probability
measure of a complete observation $\{X^k,\Phi_k\}$ of a generic sample path $%
X^k$ and by $\mathbb{E}_{\theta}$ the expectation under $\mathbb{P}_{\theta}$.

The observed likelihood function $L^{k}(\theta ) $ of $X^{k},1\leq k\leq K,$
is 
\begin{eqnarray}
L^{k}(\theta ) &=& \sum_{m=1}^M \mathbb{P}_{\theta}(X^k, R_k=m) =
\sum_{m=1}^{M}\prod_{i\in E}\mathbb{P}_{%
\theta}(X^{k}-,X_{0}^{k}=i,R_{k}=m)^{B_{i}^{k}}  \notag \\
&=&\sum_{m=1}^{M}\prod_{i\in E}\mathbb{P}_{%
\theta}(X_{0}^{k}=i)^{^{B_{i}^{k}}}\mathbb{P}_{%
\theta}(R_{k}=m|X_{0}^{k}=i)^{B_{i}^{k}}\mathbb{P}_{%
\theta}(X^{k}-|X_{0}^{k}=i,R_{k}=m)^{B_{i}^{k}}  \notag \\
&=&\prod_{i\in E}\pi _{i}^{B_{i}^{k}}\sum_{m=1}^{M}\left\{
\prod_{i\in E}\phi _{i,m}^{B_{i}^{k}}\prod_{i\in E}\left(
\prod_{j\neq i,j\in S}(q_{ij,m})^{N_{ij}^{k}}\exp\Big[\Big(%
-\sum_{j\neq i,j\in S}q_{ij,m}\Big)T_{i}^{k}\Big]\right) \right\}  \label{a1}
\end{eqnarray}%
where $\pi _{i}=\mathbb{P}_{\theta}(X_{0}=i)$ with $\sum_{i\in E}\pi _{i}=1,$
is an initial distribution of the mixture. The loglikelihood function of the 
$K$ observed sample paths $\mathcal{D}=\bigcup_{k=1}^K X^k$ is 
\begin{eqnarray}
\log L(\theta ) &=&\sum_{k=1}^{K}\log L^{k}(\theta
)=\sum_{k=1}^{K}\sum_{i\in E}B_{i}^{k}\log \pi _{i}  \notag \\
&+&\sum_{k=1}^{K}\log \sum_{m=1}^{M}\left\{ \prod_{i\in E}\phi
_{i,m}^{B_{i}^{k}}\prod_{i\in E}\left( \prod_{j\neq i,j\in
S}(q_{ij,m})^{N_{ij}^{k}}\exp \Big[\Big(-\sum_{j\neq i,j\in S}q_{ij,m}\Big)%
T_{i}^{k}\Big]\right) \right\}  \label{a2}
\end{eqnarray}%
where we can rewrite the first term as $\sum_{i\in
E}\sum_{k=1}^{K}B_{i}^{k}\log \pi _{i}=\sum_{i\in E}B_{i}\log \pi _{i},$ to
see that, up to a constant, it corresponds to the loglikelihood of the
multinomial distribution with parameters $K=\sum_{i\in E}B_{i}$ and $\pi
=(\pi _{i},i\in E),$ where $B_{i},i\in E$ are multinomial random variables$%
.\ $Hence, as is well known, the MLE of $\pi _{i}$ is $\widehat{\pi }%
_{i}=B_{i}/K.$ Therefore, when writing the complete loglikelihood function
below, we will omit the term involving $\pi .$ We also note that the
likelihood function (\ref{a2}) is the same when either the end-of-study $T$
is a fixed time in which case the last state occupation time may be right
censored, or $T$ is the first exit time to an absorbing state of the mixture
process.

It is in general difficult to obtain the MLE $\widehat{\theta }$ by directly
maximizing the observed loglikelihood function (\ref{a2}). To obtain $\widehat{\theta }$, we use the EM algorithm, which requires the loglikelihood function under complete information derived below.

\subsection{The loglikelihood function under complete information}

We now assume that we have complete information $\bigcup_{k=1}^{K}%
\{X^{k},R_{k}\}$, that is, we also know $\Phi _{k,m}(1\leq k\leq K,1\leq
m\leq M).$ By (\ref{a2}), the complete loglikelihood of the k'th realization 
$X^{k}$ is 
\begin{equation*}
\log L_{c}^{k}(\theta )=\log L_{c}^{k}(\phi )+\log L_{c}^{k}(q),
\end{equation*}%
where 
\begin{equation}
\log L_{c}^{k}(\phi )=\sum_{m=1}^{M}\Phi _{k,m}\sum_{i\in E}B_{i}^{k}\log
\phi _{i,m},  \notag
\end{equation}%
and since $\sum_{m=1}^{M}\phi _{i,m}=1,$ it depends only on $\phi =(\phi
_{i,m},i\in E,1\leq m\leq M-1).$ Therefore we can express it
as 
\begin{equation}
\log L_{c}^{k}(\phi )=\sum_{m=1}^{M-1}\Phi _{k,m}\sum_{i\in E}B_{i}^{k}\log
(\phi _{i,m})+\Phi _{k,M}\sum_{i\in E}B_{i}^{k}\log \Big(1-\sum_{m=1}^{M-1}%
\phi _{i,m}\Big).  \label{lkfi}
\end{equation}%
Now%
\begin{equation}
\log L_{c}^{k}(q)=\sum_{m=1}^{M}\Phi _{k,m}\sum_{i\in E}\sum_{j\neq i,j\in S}%
\Big(N_{ij}^{k}\log q_{ij,m}-q_{ij,m}T_{i}^{k}\Big)  \label{lkq}
\end{equation}%
and depends only on the intensities $q=\{q_{ij,m},(i,j),i\in E,j\in S,1\leq
m\leq M\}.$ Then the full information loglikelihood is 
\begin{eqnarray}
\log L_{c}(\theta ) &\sim &\sum_{k=1}^{K}\log L_{c}^{k}(\theta
)=\sum_{k=1}^{K}\left[ \log L_{c}^{k}(\phi )+\log L_{c}^{k}(q)\right]  
\notag \\
&=&\sum_{m=1}^{M}\sum_{i\in E}\Big[\sum_{k=1}^{K}\Phi _{k,m}B_{i}^{k}\log
\phi _{i,m}+\sum_{k=1}^{K}\Phi _{k,m}\sum_{j\neq i,j\in S}\Big(%
N_{ij}^{k}\log q_{ij,m}-q_{ij,m}T_{i}^{k}\Big)\Big]  \notag \\
&=&\sum_{m=1}^{M}\sum_{i\in E}B_{i,m}\log \phi
_{i,m}+\sum_{m=1}^{M}\sum_{i\in E}\sum_{j\neq i,j\in S}\Big(N_{ij,m}\log
q_{ij,m}-q_{ij,m}T_{i,m}\Big),  \label{c2}
\end{eqnarray}%
\newline
where the "wiggle" after $\log L_{c}(\theta )$ signifies that we omitted the
part of $\log L_{c}(\theta )$ which involves $\pi ,$ $B_{i,m}=\sum_{k=1}^{K}%
\Phi _{k,m}B_{i}^{k}=$the number of\ regime $m$ realizations with initial
state $i,N_{ij,m}=\sum_{k=1}^{K}\Phi _{k,m}N_{ij}^{k}=$the total number of $%
i\rightarrow j,(i\in E,j\in S)$ transitions and $T_{i,m}=\sum_{k=1}^{K}\Phi
_{k,m}T_{i}^{k}=$the total waiting time in state $i\in E$ for regime $m$
realizations.

In the sequel, we will consider the following score functions. By (\ref{lkfi}%
), the score function of the k'th complete observation with respect to $\phi
_{i,m}$ is 
\begin{equation}
\frac{\partial \log L_{c}^{k}(\phi )}{\partial \phi _{i,m}}=\frac{\Phi
_{k,m}B_{i}^{k}}{\phi _{i,m}}-\frac{\Phi _{k,M}B_i^k}{\phi _{i,M}},
\label{skfi}
\end{equation}%
and by (\ref{lkq}), the similar score function with respect to $q_{ij,m}$ is 
\begin{equation}
\frac{\partial \log L_{c}^{k}(q)}{\partial q_{ij,m}}=\frac{\Phi
_{k,m}N_{ij}^{k}}{q_{ij,m}}-\Phi _{k,m}T_{i}^{k}.  \label{skq}
\end{equation}%
Finally, the score function for all complete observations with respect to $%
\phi _{i,m}$ is 
\begin{equation}
\frac{\partial \log L_{c}(\phi )}{\partial \phi _{i,m}}=\frac{B_{i,m}}{\phi
_{i,m}}-\frac{B_{i,M}}{\phi _{i,M}},\text{ $m=1,\cdots,M-1$}.
\label{sfi}
\end{equation}%
and with respect to $q_{ij,m}$ is 
\begin{equation}
\frac{\partial \log L_{c}(\phi )}{\partial q_{ij,m}}=\frac{N_{ij,m}}{q_{ij,m}%
}-T_{i,m},\text{ $m=1,\cdots ,M.$}  \label{sq}
\end{equation}

\section{Maximum likelihood estimation of the parameter $\theta$}

The following result relates the score function of $\bigcup_{k=1}^{K}%
\{X^{k},R_{k}\}$ and that of incomplete information $\mathcal{D}%
=\bigcup_{k=1}^{K}X^{k}$. It is used to derive the MLE $\widehat{\theta }$,
its asymptotic properties and, in Section \ref{sec:sec4}, an explicit form
of the observed Fisher information matrix $I(\theta )=-\partial ^{2}\log
L(\theta )/\partial \theta ^{2}$.

\begin{lemma}
\label{lem:identity} For any $\theta \in \Theta $, the
incomplete-information score function of $\mathcal{D}$ is given by 
\begin{equation*}
\frac{\partial \log L(\theta )}{\partial \theta }=\mathbb{E}_{\theta }\Big[%
\frac{\partial \log L_{c}(\theta )}{\partial \theta }\Big\vert\mathcal{D}%
\Big].
\end{equation*}
\end{lemma}

\begin{proof}
By the first line in (\ref{c2}), 
\begin{eqnarray*}
\mathbb{E}_{\theta }\left[ \frac{\partial \log L_{c}(\theta )}{\partial
\theta }|\mathcal{D}\right]  &=&\sum_{k=1}^{K}\mathbb{E}_{\theta }\Big[\frac{%
\partial \log L_{c}^{k}(\theta )}{\partial \theta }\Big\vert X^{k}\Big] \\
&=&\sum_{k=1}^{K}\mathbb{E}_{\theta }\Big[\frac{\partial \log \mathbb{P}%
_{\theta }(X^{k},R_{k})}{\partial \theta }\Big\vert X^{k}\Big] \\
&=&\sum_{k=1}^{K}\sum_{m=1}^{M}\frac{\partial \log \mathbb{P}_{\theta
}(X^{k},R_{k}=m)}{\partial \theta }\mathbb{P}_{\theta }(R_{k}=m\big\vert %
X^{k}\big) \\
&=&\sum_{k=1}^{K}\frac{1}{\mathbb{P}_{\theta }(X^{k})}\sum_{m=1}^{M}\frac{%
\partial }{\partial \theta }\mathbb{P}_{\theta }(X^{k},R_{k}=m),
\end{eqnarray*}%
where the last equality follows from applying the Bayes formula for conditional probability. Noting that $\mathbb{P}_{\theta}(X^{k})=\sum_{m=1}^{M}\mathbb{P}_{\theta }(X^{k},R_{k}=m)$ completes the
proof.
\end{proof}

\medskip

Define $\widehat{B}_{i,m}(\theta)=\mathbb{E}_{\theta }\big[B_{i,m}%
\big\vert 
\mathcal{D}\big]$, $\widehat{N}_{ij,m}(\theta)=\mathbb{E}_{\theta }\big[%
N_{ij,m}\big\vert \mathcal{D}\big]$ and $\widehat{T}_{i,m}(\theta)=\mathbb{E}%
_{\theta }\big[T_{i,m}\big\vert \mathcal{D}\big]$. The MLE $\widehat{\theta}$
as the solution of the following systems of equations is presented below 
\begin{equation*}
\frac{\partial \log L(\theta)}{\partial \theta}=0.
\end{equation*}

\begin{proposition}
\label{lem:lem1} For $1\leq m\leq M,$ the maximum likelihood estimates of $%
\phi_{i,m}$ and $q_{ij,m}$ are 
\begin{eqnarray}
\widehat{\phi}_{i,m} &=&\frac{\widehat{B}_{i,m}(\widehat{\theta})}{B_i}\;%
\text{ }\left( i\in E\right) ,  \label{1} \\
\widehat{q}_{ij,m}&=&\frac{\widehat{N}_{ij,m}(\widehat{\theta})}{\widehat{T}%
_{i,m}(\widehat{\theta})}\;\text{ }(j\neq i,i\in E,j\in S).  \label{2}
\end{eqnarray}
\end{proposition}

\begin{proof}
Using the score function of $\phi _{i,m}$ in (\ref{sfi}) and Lemma \ref%
{lem:identity}, the MLE $\widehat{\phi }_{i,m}$ is obtained by setting $%
\mathbb{E}_{\theta }\big[\frac{\partial \log L_{c}(\phi )}{\partial \phi
_{i,m}}\big\vert\mathcal{D}\big]=0,$ which yields $\widehat{\phi }_{i,m}=%
\frac{\widehat{B}_{i,m}(\widehat{\theta })}{\widehat{B}_{i,M}(\widehat{%
\theta })}\widehat{\phi }_{i,M}$. Using the constraint $\sum_{m=1}^{M}%
\widehat{\phi }_{i,m}=1$ and $\sum_{m=1}^{M}\widehat{B}_{i,m}(\theta )=B_{i}$
for any $\theta \in \Theta $, we have $\widehat{\phi }_{i,M}=\frac{\widehat{B%
}_{i,M}(\widehat{\theta })}{B_{i}}$ and $\widehat{\phi }_{i,m}=\frac{%
\widehat{B}_{i,m}(\widehat{\theta })}{B_{i}}$. The estimator $\widehat{q}%
_{ij,m}$ is obtained by using the score function for $q_{ij,m}$ from (\ref%
{sq}) and setting the equation $\mathbb{E}_{\theta }(\frac{\partial \log
L_{c}(q)}{\partial q_{ij,m}}|\mathcal{D)}=0.$
\end{proof}

\medskip

In the next section, the EM algorithm is employed to find $\widehat{\phi }%
_{i,m}$ and $\widehat{q}_{ij,m}$. The results (\ref{1})-(\ref{2}) will be
used in the M-step of the EM-algorithm.

\subsection{The EM algorithm}

\label{sec:EM} We now use the EM algorithm to obtain the MLE of $\theta
=(\theta _{m},1\leq m\leq M)$ from $K$ realizations. Following Dempster 
\textit{et al}. (1977), the MLE of $\theta $ is found iteratively so that
the $(p+1)-$th estimate $\theta ^{p+1}$ of $\theta $ maximizes the random
function 
\begin{equation*}
\mathbb{E}_{\theta ^{p}}\big[\log L_{c}(\theta )\big|\mathcal{D}\big],
\end{equation*}%
where $\mathbb{E}_{\theta ^{p}}[\bullet |\mathcal{D}]$ refers to the
conditional expectation evaluated using the current estimate $\theta ^{p}$
after $p$ steps of the algorithm. This is the maximization step (\textit{%
M-step}) of the algorithm. The evaluation of the above conditional
expectation forms the \textit{E-step.}

\begin{proposition}
\label{prop:main} The EM algorithm for the finite g-mixture of Markov jump
processes $X=\{X_{m},1\leq m\leq M\}$ based on $K$ realizations for a fixed $M>1$ goes as follows.

\medskip

\textbf{Step 1} (\textbf{Initial step}) For $1\leq m\leq M,$ choose initial
value $\theta _{m}^{0}\equiv (q_{ij,m}^{0},i\neq j,\phi _{i,m}^{0},i\in
E,j\in S)$ of $\theta _{m}.$

\textbf{Step 2 (E-step)} At the p'th iteration $(p=0,1,2...)$, for $1\leq
m\leq M,i\in E,1\leq k\leq K,$ compute the probability of observing $X^{k}$
under $\theta _{m}^{p}$ 
\begin{eqnarray*}
L^{k}(\theta _{m}^{p})\sim \phi _{i_{k},m}^{p}\prod_{i\in E}\left\{
\prod_{j\neq i,j\in S}(q_{ij,m}^{p})^{N_{ij}^{k}}\,{\exp [(-\sum
q_{ij,m}^{p})T_{i}^{k}]}\right\} ,1\leq m\leq M.
\end{eqnarray*}%
and then the probability that the path $X^{k}$ comes from regime $m$%
\begin{eqnarray*}
\mathbb{E}_{\theta^{p}}[\Phi _{k,m}|X^{k}]=\frac{L^{k}(\theta _{m}^{p})}{%
\sum_{\ell =1}^{M}L^{k}(\theta _{\ell }^{p})}.
\end{eqnarray*}

For $i\in E,j\in S,$ and $1\leq m\leq M,$ compute%
\begin{eqnarray*}
\mathbb{E}_{\theta^{p}}[N_{ij,m}|\mathcal{D}] &\equiv
&\sum_{k=1}^{K}N_{ij}^{k}\mathbb{E}_{\theta^{p}}[\Phi
_{k,m}|X^{k}],\;\;j\neq i \\
\mathbb{E}_{\theta^{p}}[T_{i,m}|\mathcal{D}] &\equiv &\sum_{k=1}^{K}T_{i}^{k}%
\mathbb{E}_{\theta^{p}}[\Phi _{k,m}|X^{k}], \\
\mathbb{E}_{\theta^{p}}[B_{i,m}|\mathcal{D}] &\equiv &\sum_{k=1}^{K}B_{i}^{k}%
\mathbb{E}_{\theta^{p}}[\Phi _{k,m}|X^{k}].
\end{eqnarray*}

\textbf{Step 3 (M-step)} For $1\leq m\leq M,$ compute, using (\ref{1}) and (%
\ref{2}), the values $\theta _{m}^{p+1}\equiv (q_{ij,m}^{p+1},i\neq j,$ $%
\phi _{i,m}^{p+1},i\in E,j\in S)$ by 
\begin{eqnarray}
\phi _{i,m}^{p+1} &=&\frac{\mathbb{E}_{\theta^{p}}[B_{i,m}|\mathcal{D}]}{%
B_{i}},  \label{3d} \\
q_{ij,m}^{p+1} &=&\frac{\mathbb{E}_{\theta^{p}}[N_{ij,m}|\mathcal{D}]}{%
\mathbb{E}_{\theta^{p}}[T_{i,m}|\mathcal{D}]}.  \label{3c}
\end{eqnarray}

\textbf{Step 4} Stop if convergence criterion is achieved, that is, if the
Euclidean norm $||\theta ^{p+1}-\theta ^{p}||<\varepsilon ,$ where $%
\varepsilon $ is a desired small value. Otherwise, return to step 2 and
replace $\theta _{m}^{p}$ by $\theta _{m}^{p+1}.$

\medskip
\end{proposition}

For the choice of mixture model, repeat the EM algorithm for
different values of $M>1$ until the fitted model has the lowest Akaike
information criterion (AIC), see Akaike (1973). This method is applied in
Section \ref{sec:appl} to the ventICU dataset.


The EM algorithm described above converges to $\widehat{\theta }_{m},$ the
MLE of $\theta _{m}.$ To be more precise, at convergence, the equations in (%
\ref{3d}) and (\ref{3c}) take the form (\ref{1}) and (\ref{2}) respectively,
that is 
\begin{eqnarray*}
\widehat{\phi }_{i,m} &=&\frac{\mathbb{E}_{\widehat{\theta}}[B_{i,m}|%
\mathcal{D}]}{B_{i}}\;\;=\;\;\frac{\widehat{B}_{i,m}(\widehat{\theta})}{B_i},
\\
\widehat{q}_{ij,m} &=&\frac{\mathbb{E}_{\widehat{\theta}}[N_{ij,m}|\mathcal{D%
}]}{\mathbb{E}_{\widehat{\theta}}[T_{i,m}|\mathcal{D}]}\;\;=\;\;\frac{%
\widehat{N}_{ij,m}(\widehat{\theta})}{\widehat{T}_{i,m}(\widehat{\theta})}.
\end{eqnarray*}%
%
%
%
%
%
%
%
%

\subsection*{Initialization of the EM algorithm}

\label{subsec:EMinitial}

Suppose we have $K$\ sample paths from the mixture with $M$ regimes obtained from the simulation or real data. For the initial distribution $\pi _{i},i\in E$\ of the g-mixture$,$ we use the estimate $%
\widehat{\pi }_{i}$. For each $i\in E,$ we set the initial value of ($\phi
_{i,m}$, $m\in M)$ to be a uniform distribution: ($\phi _{i,m}^{0},m\in
M)=(1/M,\cdots ,1/M).$ The initial value $Q_{m}^{0}$ for the intensity
matrix $Q_{m}$ is obtained by first randomly dividing $K$ sample paths into $%
M$ subsets with the first $(M-1)$ subsets having size $\lfloor K/M\rfloor $
each, and the $M-$th subset being of size $K-(M-1)\lfloor K/M\rfloor $.
Then, treating the $m$'th subset, $m\in M,$ as if it contained realizations
from a single Markov process $X_{m}$ with intensity matrix $Q_{m},$ $%
Q_{m}^{0}\ $was set to be the MLE of the intensity matrix $Q_{m}.$

\subsection{Consistency and asymptotic normality of the MLEs}

Below we state the results about consistency and asymptotic distribution of
the MLE $\widehat{\theta }$ when the number of sample paths $K$ increases.
The results are valid for a fixed $T$ and $T$ being the absorption time.
They generalize the results from Theorem 6.1 in Albert (1962) about the
asymptotic properties of the MLE of an intensity matrix of a single Markov
process. Due to the presence of the term $\widehat{\Phi }_{k,m}=E_{\theta
^{0}}[\Phi_{k,m}\vert X^{k}]$ in the estimator $\widehat{\theta },$ the
proofs require a different approach compared to the simpler proofs of the
analogous results in Theorem 6.1. For example, consider the estimator $%
\widehat{q}_{ij,m}=\sum_{k=1}^{K}$ $\widehat{\Phi }_{k,m}N_{ij}^{k}/%
\sum_{k=1}^{K}$ $\widehat{\Phi }_{k,m}T_{i}^{k}$ of $q_{ij,m}.$ By iid
property of all K realizations $\{X^{k}\}$, which were generated under the
probability measure $\emph{P}_{\theta ^{0}}$, $\widehat{q}_{ij,m}$ converges
by LLN as $K\rightarrow \infty $ to $E_{\theta ^{0}}\big[E_{\widehat{\theta }%
}[\Phi _{k,m}N_{ij}^{k}|X^{k}]\big]/E_{\theta ^{0}}\big[E_{\widehat{\theta }%
}[\Phi _{k,m}T_{i}^{k}|X^{k}]\big].$ However, the law of iterated
expectation does not simplify the latter to $E_{\theta ^{0}}\big[\Phi
_{k,m}N_{ij}^{k}\big]/E_{\theta ^{0}}\big[\Phi _{k,m}T_{i}^{k}\big],$ which
by Lemma A.1 in SM is equal to $q_{ij,m}^{0}.$ But, for the estimator $\widehat{q}%
_{ij}$=$\sum_{k=1}^{K}$ $N_{ij}^{k}/\sum_{k=1}^{K}T_{i}^{k}$ of $q_{ij}$
considered in Albert, it immediately follows by LLN that $\widehat{q}_{ij}$
converges to $E_{\theta ^{0}}[N_{ij}^{k}]/E_{\theta ^{0}}[T_{i}^{k}]$ as $%
K\rightarrow \infty .$

To establish consistency of $\widehat{\theta }$, we use the
Shanon-Kolmogorov information inequality, see p. 113 in Ferguson (1996): for
any $\theta \neq \theta ^{0}$ and generic paths $X^{k}$, 
\begin{equation*}
R(\theta ^{0},\theta ):=\mathbb{E}_{\theta ^{0}}\Big[\log \Big(\frac{%
L^{k}(\theta ^{0})}{L^{k}(\theta )}\Big)\Big]>0,
\end{equation*}%
which in turn implies that $\theta ^{0}$ is the global maximum of $M(\theta
):=\mathbb{E}_{\theta ^{0}}\big[\log L^{k}(\theta )\big]$.

\begin{proposition}[Consistency of $\widehat{\theta}$]
By independence of $\{X^k\}$, $\widehat{\theta}\overset{\mathbb{P}_{\theta^0}%
}{\Longrightarrow}\theta^0$.
\end{proposition}

\begin{proof}
We have assumed that the parameter space $\Theta$ is a compact set of any positive values of $\theta$. The MLE $\widehat{\theta}$ is defined as the global maximizer of the sample loglikelihood $M_K(\theta)=%
\frac{1}{K}\sum_{k=1}^K \log L^k(\theta)$. By independence of $\{X^k\}$ and
the law of large numbers, $M_K(\theta)\overset{\mathbb{P}_{\theta^0}}{%
\Longrightarrow} M(\theta)=\mathbb{E}_{\theta^0}\big[\log L(\theta)\big]$.
Meanwhile, from the Shanon-Kolmogorov information inequality we have $%
\sup_{\theta \in \Theta\backslash \theta^0} M(\theta) < M(\theta^0)
\iff\theta^0=\text{arg}\max_{\theta\in\Theta} M(\theta).$ Since $%
\widehat{\theta}$ is the global maximizer of $M_K(\theta)$ and the latter
converges with probability one to $M(\theta)$, it follows that $\widehat{%
\theta}$ gets closer and closer to the global maximizer $\theta^0$ of $%
M(\theta)$ as the sample size $K$ increases, which implies that $\widehat{%
\theta} \overset{\mathbb{P}_{\theta^0}}{\Longrightarrow} \theta^0$.
\end{proof}

%

\medskip

To establish asymptotic property of the MLE $\widehat{\theta}$, the
following result is required.

\begin{lemma}
\label{lem:lem2} For any $\theta\in \Theta$,
\begin{equation*}
\mathbb{E}_{\theta }\Big[\frac{\partial \log L^{k}(\phi )}{\partial \phi
_{i,m}}\Big]=0\quad \text{and}\quad \mathbb{E}_{\theta }\Big[\frac{\partial
\log L^{k}(q)}{\partial q_{ij,m}}\Big]=0.
\end{equation*}
\end{lemma}

\begin{proof}
From (\ref{skfi}) and $\mathbb{E}_{\theta }\big[\Phi _{k,m}B_{i}^{k}\big]%
=\phi _{i,m}\pi _{i}$ for $i\in E$, $1\leq m\leq M,$ we have by Lemma \ref%
{lem:identity}, $\mathbb{E}_{\theta }\big[\frac{\partial \log L^{k}(\phi )}{%
\partial \phi _{i,m}}\big]=\mathbb{E}_{\theta }\big[\mathbb{E}_{\theta }\big[%
\frac{\partial \log L_{c}^{k}(\phi )}{\partial \phi _{i,m}}\big\vert\mathcal{%
D}\big]\big]$ =0, which gives the first result. Similarly, by (\ref{skq})
and Lemma A.1 in SM, $\mathbb{E}_{\theta }\big[\Phi _{k,m}N_{ij}^{k}\big]%
=q_{ij,m}\int_{0}^{T}\pi ^{\top }D_{m}e^{Q_{m}u}e_{i}du$ and $\mathbb{E}%
_{\theta }\big[\Phi _{k,m}T_{i}^{k}\big]=\int_{0}^{T}\pi ^{\top
}D_{m}e^{Q_{m}u}e_{i}du$, implying that $\mathbb{E}_{\theta }\big[\frac{%
\partial \log L_{c}^{k}(\phi )}{\partial q_{i,j,m}}\big]=0,$ which is the
second result.
\end{proof}

\begin{theorem}
\label{theo:theo1} As the sample size $K$ increases, $\sqrt{K}\big(\widehat{%
\theta }-\theta ^{0}\big)\overset{d}{\Longrightarrow }N\big(0,J^{-1}(\theta
^{0})\big)$ with 
\begin{equation*}
J(\theta ^{0})=\mathbb{E}_{\theta }\Big[-\frac{\partial ^{2}\log L(\theta )}{%
\partial \theta ^{2}}\Big],
\end{equation*}%
being Fisher information matrix.
\end{theorem}

\begin{proof}
From Lemma \ref{lem:identity}, it is readily verified that the
loglikelihood function $\log L(\theta)$ is twice continuously differentiable. 
By the Mean Value Theorem applied to the score function $\frac{\partial
\log L(\widehat{\theta })}{\partial \theta }$ at the true value $\theta ^{0}$%
, see Feng et al. (2013) and p.20 in Ferguson (1996), we have 
\begin{eqnarray*}
0=\frac{1}{K}\frac{\partial \log L(\widehat{\theta })}{\partial \theta }= 
\frac{1}{K}\sum_{k=1}^{K}\frac{\partial \log L^{k}(\theta ^{0})}{\partial
\theta }+\left( \int_{0}^{1}\frac{1}{K}\sum_{k=1}^{K}\frac{\partial ^{2}\log
L^{k}(\theta _{0}+\lambda (\widehat{\theta }-\theta ^{0}))}{\partial \theta
^{2}}d\lambda \right) \big(\widehat{\theta }-\theta ^{0}\big).
\end{eqnarray*}%
By consistency of $\widehat{\theta }$ and the fact that $\theta ^{0}$ is
maximizer of the function $M(\theta )$ for which $M^{\prime \prime }(\theta
^{0})=\mathbb{E}_{\theta ^{0}}\Big[\frac{\partial ^{2}\log L^{k}(\theta ^{0})%
}{\partial \theta ^{2}}\Big]<0$, hence is invertible\footnote{%
on account of Theorem 7.2.1 on p. 438 of Horn and Johnson (2013) that every
positive definite matrix is invertible and the inverse itself is positive
definite.}, we have by Slutsky's and CLT theorem 
\begin{align*}
\sqrt{K}\big(\widehat{\theta }-\theta ^{0}\big)=& \left( -\int_{0}^{1}\frac{1%
}{K}\sum_{k=1}^{K}\frac{\partial ^{2}\log L^{k}(\theta _{0}+\lambda (%
\widehat{\theta }-\theta ^{0}))}{\partial \theta ^{2}}d\lambda \right) ^{-1}
\\
& \hspace{1cm}\frac{1}{\sqrt{K}}\sum_{k=1}^{K}\left( \frac{\partial \log
L^{k}(\theta ^{0})}{\partial \theta }-\mathbb{E}_{\theta ^{0}}\Big[\frac{%
\partial \log L^{k}(\theta ^{0})}{\partial \theta }\Big]\right) \overset{d}{%
\Longrightarrow }N(0,J^{-1}(\theta ^{0})),
\end{align*}%
where by Lemma, \ref{lem:lem2} $\mathbb{E}_{\theta ^{0}}\big[\frac{\partial
\log L^{k}(\theta ^{0})}{\partial \theta }\big]=0$, which completes the
proof.
\end{proof}

\medskip

Following the above theorem, estimates of standard errors of the MLE $%
\widehat{\theta }$ are calculated in Section \ref{sec:sec4} using the
inverse of the observed Fisher information $I(\theta)$.

However, it is difficult to derive an explicit form of the asymptotic
information matrix $J(\theta)$, see Theorem \ref{theo:Fisher} in Section \ref%
{sec:sec4}. Instead, in the following section we derive a
lower bound for the asymptotic matrix.

\subsection{Lower bound for the asymptotic variance of the MLEs}

To state the corresponding lower bound for asymptotic variance of 
$\widehat{\theta }$, for $1\leq n\leq M-1,$ let $D_{n}$ be a ($w\times w$) diagonal
matrix with diagonal elements $\phi _{i,n},i\in E,$ and $e_{i}$ $%
=(0,...1,...0)$ a $(w\times 1$) unit vector with value one on the ith
component and zero otherwise. Denote by $\theta _{0}=(\phi ^{0},q^{0})$ the
true value of $\theta =(\phi ,q)$ where $\phi =(\phi _{i,m},i\in E,1\leq
m\leq M-1)\ $and\ $q=(q_{ij,m}$,$j\neq i,i\in E,j\in S,1\leq m\leq M),$ and
define an indicator function%
\begin{equation*}
\delta _{q}(z)=\left\{ 
\begin{array}{lr}
1, & q=z \\[6pt] 
0, & \text{otherwise}.%
\end{array}%
\right.
\end{equation*}

\begin{theorem}
\label{theo:main} Let $T$ be either a fixed time or the absorption time.
Then, as $K\rightarrow \infty ,$ 
\begin{equation}  \label{eq:infoloss}
J^{-1}(\theta^0)\;\geq\; \Sigma(\theta^0),
\end{equation}%
where $\Sigma(\theta^0) :=\text{Cov}(\widehat{\theta}-\theta^0,\widehat{%
\theta } _{0}-\theta^0)$ is a $(w(Mw-1)\times w(Mw-1))$ block-diagonal matrix
with 
\begin{equation}  \label{6a}
\Sigma(\theta^0)= 
\begin{cases}
\text{Cov}(\widehat{\phi }_{r,n}-\phi _{r,n}^{0},\widehat{\phi }_{i,m}-\phi
_{i,m}^{0}) & =\;\;\frac{\phi _{r,m}^{0}}{\pi _{r}}\delta _{i}(r)(\delta
_{m}(n)-\phi _{r,n}^{0}), \\[8pt] 
\text{Cov}(\widehat{\phi }_{r,n}-\phi _{r,n}^{0},\widehat{q}%
_{ij,m}-q_{ij,m}^{0}) & =\;\;0 \\[8pt] 
\text{Cov}(\widehat{q}_{rv,n}-q_{rv,n}^{0},\widehat{q}_{ij,m}-q_{ij,m}^{0})
& =\;\;\frac{q_{rv,n}^{0}\delta _{m}(n)\delta _{i}(r)\delta _{j}(v)}{\mathbb{%
E}_{\theta^0}(\Phi _{k,n}T_{i}^{k})},%
\end{cases}%
\end{equation}%
where for $D_{n}^{0}=\text{diag}(\phi _{1,n}^{0},\cdots ,\phi _{w,n}^{0})$, $%
Q_{n}^{0}=[q_{ij,n}^{0}]_{ij}$, and $\pi =(\pi _{1},\cdots ,\pi _{w})$, 
\begin{equation*}
\mathbb{E}_{\theta^0}[\Phi _{k,n}T_{i}^{k}]=\Bigg\{%
\begin{array}{lr}
\int_{0}^{T}\pi ^{\top }D_{n}^{0}e^{uQ_{n}^{0}}e_{i}du, & \text{for fixed $T$%
} \\[8pt] 
\int_{0}^{\infty }\pi ^{\top }D_{n}^{0}e^{uQ_{n}^{0}}e_{i}du, & \text{for
absorption time $T$}.%
\end{array}%
\end{equation*}
\end{theorem}

\begin{proof}
See part A in the Supplementary material.
\end{proof}

\medskip

The inequality (\ref{eq:infoloss}) corresponds to the resulting information
loss presented in incomplete data, see Schervish (1995). In the absence of
heterogeneity ($\delta _{m}(n)=1$ and $D_{m}=I$), hence for the case of
complete information, the result in (\ref{6a}) coincides with that of
Theorem 6.1 in Albert (1962). In the presence of heterogeneity with complete
information, estimators of transition rates for different regimes have zero
covariances. There is also zero covariance between estimators of regime
memberships across different states and transition rates.

\section{The finite sample covariance matrix for the MLEs}

\label{sec:sec4}

We will use the following vectors for the computation of the estimated
variances of $\widehat{\phi }$ and $\widehat{q}$ and covariances. Define 
\begin{equation*}
\phi _{m}=(\phi _{1,m},\cdots ,\phi _{w,m}),\;\;1\leq m\leq M-1.
\end{equation*}%
We form the row vector $\phi $ of dimension $(1\times (M-1)w)$, which combines the vectors $\phi _{m}$, i.e., $\phi
=\left( \phi _{1},\cdots ,\phi _{M-1}\right) $. Similarly, let 
$q_{m}$ be a $(1\times w(w-1))-$vector formed by combining the $E$-row
vectors of $Q_{m}$, with diagonal element $q_{ii,m}$ removed, i.e., 
\begin{equation*}
q_{m}=\left( q_{12,m},\cdots ,q_{1w,m},q_{21,m},q_{23,m},\cdots
,q_{2w,m},\cdots ,q_{w1,m},\cdots ,q_{w(w-1),m}\right),
\end{equation*}%
for $1\leq m\leq M$. Next, we form a row vector $q$ of
dimension $(1\times Mw(w-1))$, when there are no absorbing states, which
combines the vectors $q_{m}$, i.e., $q=\left( q_{1},\cdots ,q_{M}\right) $.
The g-mixture parameters to be estimated are defined by the $(1\times w(Mw-1))-$vector $\theta =\left( \phi ,q\right) .$

\subsection{Observed Fisher information matrix}

Based on continuous observation of the sample paths $\mathcal{D}%
=\bigcup_{k=1}^K X^k$, the observed Fisher information matrix $I(\theta):=-%
\frac{\partial^2 \log L(\theta)}{\partial \theta^2}$ can be derived using
the identity of Lemma \ref{lem:identity}.

\begin{theorem}
\label{theo:Fisher} For any $(\theta_i,\theta_j)\in \theta$, the $(i,j)$
component $-\frac{\partial^2 \log L(\theta)}{\partial \theta_i \partial
\theta_j}$ of $I(\theta)$ is given by 
\begin{eqnarray}  \label{eq:InfMatr}
I(\theta_i,\theta_j)&=&\sum_{k=1}^K \mathbb{E}_{\theta}\Big[-\frac{%
\partial^2 \log L_c^k(\theta)}{\partial \theta_i \partial \theta_j} %
\Big\vert X^k\Big] -\sum_{k=1}^K \mathbb{E}_{\theta}\Big[ \Big(\frac{%
\partial \log L_c^k(\theta)}{\partial \theta_i}\Big)\Big( \frac{\partial
\log L_c^k(\theta)}{\partial \theta_j}\Big)\Big\vert X^k\Big]  \notag \\
&&+\sum_{k=1}^K\mathbb{E}_{\theta}\Big[ \frac{\partial \log L_c^k(\theta)}{%
\partial \theta_i}\Big\vert X^k\Big]\mathbb{E}_{\theta}\Big[ \frac{\partial
\log L_c^k(\theta)}{\partial \theta_j}\Big\vert X^k\Big].
\end{eqnarray}
\end{theorem}

Notice that the information matrix (\ref{eq:InfMatr}) is slightly different
from Louis (1982) general matrix formula. In its new form, the information
matrix (\ref{eq:InfMatr}) simplifies the conditional expectation of outer
product of the complete-data score function in the Louis' matrix formula.
The convergence of $K^{-1}I(\theta_i,\theta_j)$ to the respective element $%
J(\theta_i,\theta_j)$ of the Fisher information matrix $J(\theta)$ is more
immediate from (\ref{eq:InfMatr}) than from the Louis formula. The first
term in (\ref{eq:InfMatr}) is the expected complete data observed
information and the other two terms combined is the variance of complete
data score, conditional on $D$.

\medskip

\begin{proof}
Following the third equality in the proof of Lemma \ref{lem:identity}, we
have 
\begin{eqnarray*}
\frac{\partial \log L(\theta)}{\partial \theta_j} &=&\sum_{k=1}^K
\sum_{m=1}^M \Big(\frac{\partial \log \mathbb{P}_{\theta}(X^k,R_k=m) }{%
\partial \theta_j} \Big)\mathbb{P}_{\theta}(R_k=m\big\vert X^k).
\end{eqnarray*}
Differentiating with respect to variable $\theta_i$ on both sides of the
above equality gives 
\begin{eqnarray*}
-\frac{\partial^2 \log L(\theta)}{\partial \theta_i \partial \theta_j}
&=&\sum_{k=1}^K \sum_{m=1}^M \Big(-\frac{\partial^2 \log \mathbb{P}%
_{\theta}(X^k,R_k=m)}{\partial \theta_i\partial \theta_j} \Big)\mathbb{P}%
_{\theta}(R_k=m\big\vert X^k)  \notag \\
&&\hspace{0cm}- \sum_{k=1}^K \sum_{m=1}^M \Big(\frac{\partial \log \mathbb{P}%
_{\theta}(X^k,R_k=m)}{\partial \theta_j} \Big)\frac{\partial \mathbb{P}%
_{\theta}(R_k=m\big\vert X^k)}{\partial \theta_i}  \notag \\
&=&\sum_{k=1}^K \mathbb{E}_{\theta}\Big[-\frac{\partial^2 \log L_c^k(\theta)%
}{\partial \theta_i\partial \theta_j}\Big\vert X^k\Big]  \notag \\
&&\hspace{0cm} - \sum_{k=1}^K \sum_{m=1}^M \Big(\frac{\partial \log \mathbb{P%
}_{\theta}(X^k,R_k=m)}{\partial \theta_j} \Big)\frac{\partial \mathbb{P}%
_{\theta}(R_k=m\big\vert X^k)}{\partial \theta_i}.
\end{eqnarray*}
The final result is obtained after replacing the derivative $\frac{\partial 
\mathbb{P}_{\theta}(R_k=m\big\vert X^k)}{\partial \theta_i} $ by 
\begin{align*}
\mathbb{P}_{\theta}(R_k=m\big\vert X^k)\left(\frac{\partial \log \mathbb{P}%
_{\theta}(X^k,R_k=m) }{\partial\theta_i} -\mathbb{E}_{\theta}\Big[\frac{%
\partial \log L_c^k(\theta)}{\partial \theta_i}\Big\vert X^k\Big]\right),
\end{align*}
%
leading to the information matrix (\ref{eq:InfMatr}). 
\end{proof}

\medskip

Using the results of Proposition A.1 in SM, one can show for any $%
(\theta_i,\theta_j)\in\theta$, $\mathbb{E}_{\theta}\big[-\frac{\partial^2
\log L_c^k(\theta)}{\partial \theta_i \partial \theta_j} \big] =\mathbb{E}%
_{\theta}\big[ \big(\frac{\partial \log L_c^k(\theta)}{\partial \theta_i}%
\big)\big( \frac{\partial \log L_c^k(\theta)}{\partial \theta_j}\big)\big]$
which by Lemma \ref{lem:identity} and LLN leads to the convergence of $%
K^{-1}I(\theta_i,\theta_j)$ to $J(\theta_i,\theta_j)$, the $(i,j)-$element
of the Fisher information $J(\theta)$.

\medskip

Below we derive the explicit expressions for $I(\widehat{\phi }),I(\widehat{q%
})$ and $I(\widehat{q},\widehat{\phi }).$

\subsection{Elements of the matrices $I(\widehat{\phi })$, $%
I(\widehat{q})$, and $I(\widehat{\phi },\widehat{q})$}

To derive the expressions for $I(\widehat{\phi })$, $I(\widehat{q})$, and $I(%
\widehat{\phi },\widehat{q})$, we consider the score function $\frac{%
\partial \log L_c(\theta)}{\partial \theta}$ and its derivative, where we
let $\widehat{A}_{ij,m}^{k}=N_{ij}^{k}-\widehat{q}_{ij,m}T_{i}^{k}$ and $%
\widehat{\Psi}_{i,m\vert M}^k= \widehat{\Phi}_{k,m} -\frac{\widehat{\phi}%
_{i,m}}{\widehat{\phi}_{i,M}}\widehat{\Phi}_{k,M}$.

\begin{proposition}
\label{prop:prop2} From the information matrix (\ref{eq:InfMatr}%
), we have for $(i,r)\in E$, $(j,v)\in S$,

\begin{enumerate}
\item[(i)] and $n,m=1,\cdots,M-1$,
\begin{eqnarray*}
I(\widehat{\phi }_{j,n},\widehat{\phi }_{i,m})=\frac{\delta _{i}(j)}{%
\widehat{\phi }_{j,n}\widehat{\phi }_{i,m}}\sum_{k=1}^{K}\widehat{\Psi }%
_{i,m\vert M}^k\widehat{\Psi}_{j,n\vert M}^kB_{j}^{k},
\end{eqnarray*}

\item[(ii)] and for $n,m=1,\cdots,M$,
\begin{eqnarray*}
I(\widehat{q}_{rv,n},\widehat{q}_{ij,m})=\frac{\delta_i(r)\delta_j(v)%
\delta_m(n)}{\widehat{q}_{ij,m}\widehat{q}_{rv,n}}\widehat{N}_{rv,\ell} -%
\frac{1}{\widehat{q}_{rv,n}\widehat{q}_{ij,m}}\sum_{k=1}^{K}\widehat{\Phi }%
_{k,n}\left(\delta _{m}(n)-\widehat{\Phi }_{k,m}\right)\widehat{A}_{rv,n}^{k}%
\widehat{A}_{ij,m}^{k},
\end{eqnarray*}

\item[(iii)] and for $n=1,\cdots,M$; $m=1,\cdots, M-1$,
\begin{eqnarray*}
I(\widehat{\phi }_{i,m},\widehat{q}_{rv,n})=-\frac{1}{\widehat{\phi }_{i,m}%
\widehat{q}_{rv,n}}\sum_{k=1}^{K}\widehat{\Phi }_{k,n}\Big(\delta _{m}(n)- 
\frac{\widehat{\phi}_{i,m}}{\widehat{\phi}_{i,M}}\delta_M(n) -\widehat{\Psi }%
_{i,m\vert M}^k\Big)\widehat{A}_{rv,n}^{k}B_{i}^{k}.
\end{eqnarray*}
\end{enumerate}
\end{proposition}

\begin{proof}
The proofs of above results are provided in part B of Supplementary material.
\end{proof}

\medskip

Using expression (7) on p.12 in Magnus and Neudecker (2007) for inversion of
block partitioned matrix, inverting the information matrix $I(\widehat{%
\theta })$ yields estimated covariance of $\widehat{\theta }$ exhibited in
the following proposition.

\begin{proposition}
\label{prop:main2} The estimated covariance matrix of the MLEs $\widehat{%
\theta }$ is given by 
\begin{eqnarray*}
\widehat{\text{Var}}(\widehat{\theta })=\left( 
\begin{array}{cc}
\widehat{\text{Var}}(\widehat{\phi }) & \widehat{\text{Cov}}(\widehat{\phi },%
\widehat{q}) \\ 
\widehat{\text{Cov}}^{\top }(\widehat{\phi },\widehat{q}) & \widehat{\text{%
Var}}(\widehat{q})%
\end{array}%
\right) ,
\end{eqnarray*}%
where the estimated variances for $\widehat{q}$, $\widehat{\phi }$, and
their covariance are defined by 
\begin{eqnarray*}
\widehat{\text{Var}}(\widehat{q}) &=&[I(\widehat{q})-I(\widehat{q},\widehat{%
\phi })I^{-1}(\widehat{\phi })I(\widehat{\phi },\widehat{q})]^{-1}, \\[8pt]
\widehat{\text{Cov}}(\widehat{\phi },\widehat{q}) &=&-I^{-1}(\widehat{\phi }%
)I(\widehat{\phi },\widehat{q})[I(\widehat{q})-I(\widehat{q},\widehat{\phi }%
)I^{-1}(\widehat{\phi })I(\widehat{\phi },\widehat{q})]^{-1}, \\[8pt]
\widehat{\text{Var}}(\widehat{\phi }) &=&I^{-1}(\widehat{\phi })+\widehat{%
\text{Cov}}(\widehat{\phi },\widehat{q})[\widehat{\text{Var}}(\widehat{q}%
)]^{-1}\widehat{\text{Cov}}^{\top }(\widehat{\phi },\widehat{q}).
\end{eqnarray*}%
Thus, if $I(\widehat{\phi },\widehat{q})=0$, $\widehat{\text{Cov}}(\widehat{%
\phi },\widehat{q})=0$, $\widehat{\text{Var}}(\widehat{\phi })=I^{-1}(%
\widehat{\phi })$, and $\widehat{\text{Var}}(\widehat{q})=I^{-1}(\widehat{q}%
) $.
\end{proposition}

\begin{remark}
\label{rem:rem2} Note that in order to prevent having singularity in
estimating the variance of $\widehat{\theta }$ by inverting the information
matrix $I(\widehat{\theta })$, we exclude the estimators $\widehat{\phi }%
_{i,m}$ and $\widehat{q}_{ij,m}$ whose values are (very close) to zero.
\end{remark}

\subsection{Computation of $I(\widehat{\theta })$ for a
two-regime mixture model}

\label{sec:twomix}

We specialize the general results from Proposition \ref{prop:prop2} to the
case of two-regime mixture model defined on state space $S=E\cup \Delta ,$
where $E=\{1,2\}$ is a set of transient states and $\Delta =\{3,4\}$ is a
set of two absorbing states corresponding to the ventICU dataset used in the
application section with the state diagram described by Figure \ref%
{fig:mixabs}. Since{\ $\widehat{\phi }_{i,2}=1-\widehat{\phi }_{i,1},$ we
have $\text{Var}(\widehat{\phi }_{i,2})=\text{Var}(\widehat{\phi }_{i,1})$
for each $i\in E,$ and thus consider only the vector $\phi =(\phi
_{1,1},\phi _{2,1})^{\top }$.} The $I(\widehat{\phi })$ is a $(2\times 2)-$%
diagonal matrix with the entries given, using Proposition \ref{prop:prop2},
by 
\begin{eqnarray*}
I_{i1}(\widehat{\phi })=\sum_{k=1}^{K}\left(\frac{\widehat{\Psi }_{i,1\vert
2}^k}{\widehat{\phi }_{i,1}}\right)^{2}B_{i}^{k},\;\;i\in E.
\end{eqnarray*}

To simplify the presentation of the matrix $I(\widehat{q})$, we split it into
block partitioned matrix: 
\begin{eqnarray*}
I(\widehat{q})=\left( 
\begin{array}{cc}
I(\widehat{q}_{1}) & I(\widehat{q}_{1},\widehat{q}_{2}) \\ 
I(\widehat{q}_{2},\widehat{q}_{1}) & I(\widehat{q}_{2})%
\end{array}%
\right) .
\end{eqnarray*}%
Note that by the symmetry property, $I(\widehat{q}_{2},\widehat{q}%
_{1})=I^{\top }(\widehat{q}_{1},\widehat{q}_{2}).$ To write elements of the
matrix $I(\widehat{q}_{m})_{6\times 6}$, with $m=1,2$, we use the sequence $%
\widehat{q}_{m}=(\widehat{q}_{12,m},\widehat{q}_{13,m},\widehat{q}_{14,m},%
\widehat{q}_{21,m},\widehat{q}_{23,m},\widehat{q}_{24,m})$ and number its
components from 1 to 6, so that it is read as $\widehat{q}_{m}=(\widehat{q}%
_{1,m},...,\widehat{q}_{6,m}).$ Similarly, we label the $\ell -$th component
of the vectors $(\widehat{N}_{12,m},\widehat{N}_{13,m},\widehat{N}_{14,m},%
\widehat{N}_{21,m},\widehat{N}_{23,m},\widehat{N}_{24,m})$ and $(\widehat{A}%
_{12,m}^{k},\widehat{A}_{13,m}^{k},\widehat{A}_{14,m}^{k},\widehat{A}%
_{21,m}^{k},\widehat{A}_{23,m}^{k},\widehat{A}_{24,m}^{k})$ by $\widehat{N}%
_{\ell ,m}$ and $\widehat{A}_{\ell ,m}^{k}$.

Then, for $1\leq \ell \leq 6,$ the diagonal elements of the matrix $I(%
\widehat{q}_{m})$ are 
\begin{eqnarray*}
I_{\ell \ell }(\widehat{q}_{m})=\frac{\widehat{N}_{\ell ,m}}{\widehat{q}%
_{\ell ,m}^2}-\frac{1}{\widehat{q}_{\ell ,m}^2}\sum_{k=1}^{K}\widehat{\Phi }%
_{k,m}\big(1-\widehat{\Phi }_{k,m}\big)\left( \widehat{A}_{\ell
,m}^{k}\right) ^{2},
\end{eqnarray*}%
while, for $1\leq \ell \neq \nu \leq 6,$ the off-diagonal elements are 
\begin{eqnarray*}
I_{\ell \nu }(\widehat{q}_{m})=-\frac{1}{\widehat{q}_{\ell ,m}\widehat{q}%
_{\nu ,m}}\sum_{k=1}^{K}\widehat{\Phi }_{k,m}\big(1-\widehat{\Phi }_{k,m}%
\big)\widehat{A}_{\ell ,m}^{k}\widehat{A}_{\nu ,m}^{k}.
\end{eqnarray*}%
Moreover, the $(\ell ,\nu )-$element of the matrix $I(\widehat{q}_{1},%
\widehat{q}_{2})$ is 
\begin{eqnarray*}
I_{\ell \nu }(\widehat{q}_{1},\widehat{q}_{2})=\frac{1}{\widehat{q}_{\ell ,1}%
\widehat{q}_{\nu ,2}}\sum_{k=1}^{K}\widehat{\Phi }_{k,1}\widehat{\Phi }_{k,2}%
\widehat{A}_{\ell ,1}^{k}\widehat{A}_{\nu ,2}^{k}.
\end{eqnarray*}%
For convenience, we write $I(\widehat{\phi },\widehat{q})=[I(\widehat{\phi },%
\widehat{q}_{1}),I(\widehat{\phi },\widehat{q}_{2})]$, where for $m=1,2,$
each $I(\widehat{\phi },\widehat{q}_{m})$ is a $(2\times 6)-$matrix whose $%
(\ell ,\nu )-$element is 
\begin{eqnarray*}
I_{\ell \nu }(\widehat{\phi },\widehat{q}_{m})=\Bigg\{%
\begin{array}{lr}
-\frac{1}{\widehat{\phi }_{\ell ,1}\widehat{q}_{\nu ,1}}\sum_{k=1}^{K}%
\widehat{\Phi }_{k,1}(1-\widehat{\Psi }_{\ell,1\vert 2}^k)\widehat{A}_{\nu
,1}^{k}B_{\ell }^{k}, & \;m=1, \\[10pt] 
\frac{1}{\widehat{\phi }_{\ell ,1}\widehat{q}_{\nu ,2}}\sum_{k=1}^{K}%
\widehat{\Phi }_{k,1}\left(\frac{\widehat{\phi}_{\ell,1}}{\widehat{\phi}%
_{\ell,2}}+\widehat{\Psi }_{\ell,1\vert 2}^k\right)\widehat{A}_{\nu
,2}^{k}B_{\ell }^{k}, & \;m=2,%
\end{array}%
\end{eqnarray*}%
for $\ell =1,2$ and $1\leq \nu \leq 6$.

\section{Simulation Study}

\label{sec:MCSim}

We consider a mixture of two continuous-time Markov jump processes $X_{m},$ $%
m=1,2,$ with the intensity matrices $Q_{m},m=1,2$ on state space $\{1,2,3\}$%
. For the purpose of simulation, we express $Q_{m}$ as 
\begin{equation*}
Q_{m}=\text{diag(}q_{1,m},q_{2,m},q_{3,m})(P_{m}-I),m=1,2
\end{equation*}%
where $\text{diag(}q_{1,m},q_{2,m},q_{3,m})$ is the diagonal matrix with $%
q_{i,m}=-q_{ii,m}$ being the exit rate of $X_{m}$ from state $i,$ $P_{m}$
the transition matrix of a discrete time Markov chain $Z^{m}$ embedded in
the Markov process $X_{m}$ and $I$ an identity matrix.

For the true values of the mixture's parameters, we chose uniform initial
distribution $\pi =(1/3,1/3,1/3),$ and the regime 1 and 2 probabilities as $%
\phi _{1}=(\phi _{1,1},\phi _{2,1},\phi _{3,1})=(0.5,0.25,1/5)$ and $\phi
_{2}=(1,1,1)-\phi _{1},$ respectively. Furthermore, we chose the true
transition matrices of the embedded Markov chains $Z^{m},$ $m=1,2$, to be 
\begin{equation*}
P_{1}=\left( 
\begin{array}{ccc}
0 & 0.6 & 0.4 \\ 
0.5 & 0 & 0.5 \\ 
0.4 & 0.6 & 0%
\end{array}%
\right) \;\;\text{and}\;\;P_{2}=\left( 
\begin{array}{ccc}
0 & 0.8 & 0.2 \\ 
0.5 & 0 & 0.5 \\ 
0.2 & 0.8 & 0%
\end{array}%
\right)
\end{equation*}%
and the true exit rates from states as $q_{1}=(1/3,2/5,1/5)$ and $%
q_{2}=(1/2,1/4,3/4)$ for regime 1 and 2 respectively. The elements in the
true intensity matrices can be found in Table 1 below.

\subsection{Simulating a sample path of a two-regime mixture}

\label{subsec:MCSim}

We simulate a sample path of a two-regime mixture on time interval $%
(0,T)=(0,30).$ The simulation uses Sigman's (2017) method for simulating a
sample path of a discrete-time Markov chain. In the simulation, $W_{i}^{m}$
denotes a waiting time of $X_{m}$ in state $i$. And we write $W_{i}^{m}\sim
\exp (q_{i,m})$ to say that $W_{i}^{m}$ has an exponential distribution with
parameter $q_{i,m}.$ We also note that $Z_{0}^{m}=X_{m}(0)$ for $m=1,2.$

\textbf{Step 1} Draw at random an initial state $i_{0}$ from a uniform
distribution on states $1,2,3$.

\textbf{Step 2} Given $i_{0},$ draw the regime indicator $m$ from the
Bernoulli distribution with success probability equal to $\phi _{i_{0},1},$
where success corresponds to regime $Q_{1}$.

\textbf{Step 3} Set $j=1$

\textbf{Step 4} For\ a chosen $m$ in Step 2$,$ set $Z_{j-1}^{m}=i_{j-1},\ $%
simulate waiting time $W_{i_{j-1}}^{m}\sim \exp (q_{i_{j-1},m})$ and compute 
$S_{j-1}^{m}\equiv \sum_{k=0}^{j-1}W_{i_{k}}^{m}.$ Stop if $S_{j-1}^{m}>T.$
If not, go to Step 5

\textbf{Step 5 }Simulate $Z_{j}^{m}=i_{j},$ conditioning on state $i_{j-1}$ 
\begin{eqnarray*}
\text{if }i_{j-1} &=&1\text{ and }U_{j}\leq p_{12,m},\text{ set }Z_{j}^{m}=2
\\
\text{if }i_{j-1} &=&1\text{ and }U_{j}>p_{12,m},\text{ set }Z_{j}^{m}=3,
\end{eqnarray*}%
\begin{eqnarray*}
\text{if }i_{j-1} &=&2\text{ and }U_{j}\leq p_{21,m},\text{ set }Z_{j}^{m}=1
\\
\text{if }i_{j-1} &=&2\text{ and }U_{j}>p_{21,m},\text{ set }Z_{j}^{m}=3,
\end{eqnarray*}%
\begin{eqnarray*}
\text{if }i_{j-1} &=&3\text{ and }U_{1}\leq p_{31,m},\text{ set }Z_{j}^{m}=1
\\
\text{if }i_{j-1} &=&3\text{ and }U_{1}>p_{31,m},\text{ set }Z_{j}^{m}=2,
\end{eqnarray*}%
where $U_{j}$ is drawn, independently from previous draws, from $U(0,1),$ a
uniform distribution on [0,1] $.$ Increase $j$ by one and go to Step 4.

Let $J\equiv \min (j:S_{j-1}^{m}>T).$ Then $J$ is the iteration at which the
simulation stops. The resulting sample path is \{$%
Z_{0}^{m}=i_{0},W_{i_{0,}}^{m},Z_{1}^{m}=i_{1},W_{i_{1}}^{m},...,Z_{J-1}^{m}=i_{J-1},W_{i_{J-1}}^{m,C}\}, 
$ where $W_{i_{J-1}}^{m,C}$ is a right-censored waiting time in state $%
i_{J-1}$ $.$

\begin{table}[t!]
\centering 
\par
\begin{tabular}{|l|l|l|l|l|l|l|l|l|l|}
\hline
\multirow{2}{*}{$\;\;\theta$} & \multicolumn{1}{c|}{True} & 
\multicolumn{3}{c|}{Bias ($10^{-2}$) for $K=$} & \multicolumn{3}{c|}{RMSE ($%
10^{-2} $) for $K=$}   \\ 
& \,\textrm{Value} & $800$ & $1200$ & $2000$ & $800$ & $1200$ & $2000$ \\ \hline
$\phi_{1,1}$ & 0.5 & 2.0787 & 0.8372 & -0.1656 & 7.4662 & 5.1977 & 3.9595   \\ \hline
$\phi_{2,1}$ & 0.25 & 1.1391 & 0.5739 & -0.0389 & 6.8112 & 5.6301 & 3.7659   \\ \hline
$\phi_{3,1}$ & 0.75 & 0.8329 & 0.2995 & -0.2127 & 6.0929 & 5.0411 & 3.6592   \\ \hline
$q_{12,1}$ & 0.2 & 0.1579 & 0.1187 & -0.0432 & 1.1507 & 0.8344 & 0.6330 \\ \hline
$q_{13,1}$ & 0.1333 & 0.0951 & 0.0317 & 0.00003 & 0.6299 & 0.5779 & 0.4388   \\ \hline
$q_{21,1}$ & 0.2 & -0.0376 & 0.0158 & 0.0005 & 0.8900 & 0.6754 & 0.5373 \\ \hline
$q_{23,1}$ & 0.2 & -0.1477 & 0.0950 & -0.0508 & 0.8984 & 0.6626 & 0.5546   \\ \hline
$q_{31,1}$ & 0.2 & -0.1477 & -0.0661 & 0.0237 & 1.4041 & 1.1318 & 0.8727   \\ \hline
$q_{32,1}$ & 0.3 & 0.1096 & 0.0392 & -0.0225 & 1.1586 & 1.0826 & 0.6792 \\ \hline
$q_{12,2}$ & 0.4 & 0.3652 & 0.1250 & -0.0200 & 2.0643 & 1.8367 & 1.3450 \\ \hline
$q_{13,2}$ & 0.1 & -0.1462 & -0.0235 & 0.0060 & 0.8862 & 0.6501 & 0.5309   \\ \hline
$q_{21,2}$ & 0.2 & 0.0485 & 0.0394 & 0.0097 & 0.7981 & 0.6333 & 0.4349 \\ \hline
$q_{23,2}$ & 0.2 & 0.0839 & -0.0486 & -0.0058 & 0.8730 & 0.6035 & 0.4763   \\ \hline
$q_{31,2}$ & 0.0667 & -0.1083 & -0.0740 & 0.0189 & 0.7627 & 0.6022 & 0.4026   \\ \hline
$q_{32,2}$ & 0.2667 & -0.1751 & -0.1607 & 0.0580 & 1.0573 & 0.9602 & 0.6445  \\ \hline
\end{tabular}
\caption{Bias($\widehat{\theta}$)=$\frac{1}{N}%
\sum_{n=1}^N \big(\widehat{\theta}_{n,K}- \theta\big)$ and RMSE($\widehat{\theta}$) =$\Big[\frac{1}{N}%
\sum_{n=1}^N \big(\widehat{\theta}_{n,K} -\theta\big)^2\Big]^{1/2}$, with fixed $N=200$, for different number $K$ of
sample paths.}
\label{table:biasmse}
\end{table}

\begin{table}[t!]
\centering 
\par
\begin{tabular}{|l|l|l|l|l|l|l|l|l|l|l|l|}
\hline
\multirow{2}{*}{$\;\;\theta$} & \multicolumn{1}{c|}{True} & 
\multicolumn{1}{c|}{Estimate} & \multicolumn{3}{c|}{Std. Error (\%)} & 
\multicolumn{1}{c|}{}   \\ 
& \,\,\textrm{Value} & $\;\;\;\;\;\widehat{\theta}_K$ & \textrm{RMSE} & $%
\sqrt{I^{-1}(\widehat{\theta}_K)}$ & \,\,$\sqrt{\Sigma(\theta^0)}$ & \;\;%
\textrm{KS}  \\ \hline
$\phi_{1,1}$ & 0.5000 & 0.4990 & 4.1612 & 4.1559 & 1.9365 & 0.4340   \\ \hline
$\phi_{2,1}$ & 0.2500 & 0.2514 & 3.8424 & 3.8082 & 1.6771 & 0.3701   \\ \hline
$\phi_{3,1}$ & 0.7500 & 0.7484 & 3.8237 & 3.8113 & 1.6771 & 0.9209   \\ \hline
$q_{12,1}$ & 0.2000 & 0.1997 & 0.6322 & 0.6429 & 0.4231 & 0.7204   \\ \hline
$q_{13,1}$ & 0.1333 & 0.1336 & 0.4339 & 0.4153 & 0.3455 & 0.4850   \\ \hline
$q_{21,1}$ & 0.2000 & 0.2005 & 0.5357 & 0.5450 & 0.4271 & 0.1711   \\ \hline
$q_{23,1}$ & 0.2000 & 0.1999 & 0.5653 & 0.5445 & 0.4271 & 0.8198    \\ \hline
$q_{31,1}$ & 0.2000 & 0.2004 & 0.8134 & 0.8057 & 0.5043 & 0.8240   \\ \hline
$q_{32,1}$ & 0.3000 & 0.3007 & 0.7916 & 0.7717 & 0.6176 & 0.8679   \\ \hline
$q_{12,2}$ & 0.4000 & 0.3999 & 1.2999 & 1.2841 & 0.7618 & 0.5306   \\ \hline
$q_{13,2}$ & 0.1000 & 0.0996 & 0.5158 & 0.5153 & 0.3809 & 0.1062   \\ \hline
$q_{21,2}$ & 0.2000 & 0.1998 & 0.4977 & 0.4731 & 0.3854 & 0.2508   \\ \hline
$q_{23,2}$ & 0.2000 & 0.2005 & 0.4461 & 0.4744 & 0.3854 & 0.2041   \\ \hline
$q_{31,2}$ & 0.0667 & 0.0666 & 0.4181 & 0.4310 & 0.2629 & 0.8180   \\ \hline
$q_{32,2}$ & 0.2667 & 0.2670 & 0.6234 & 0.6465 & 0.5258 & 0.3770   \\ \hline
\end{tabular}
\caption{Estimated standard errors of the MLEs $\widehat{\theta}_K=\frac{1}{N}\sum_{n=1}^N \widehat{\theta}_{n,K}$
using RMSE($\widehat{\theta}_K$) =$\big[\frac{1}{N}%
\sum_{n=1}^N \big(\widehat{\theta}_{n,K} -\theta\big)^2\big]^{1/2}$ and the inverse of Fisher information $I(\widehat{\theta}_K)=\frac{1}{N}\sum_{n=1}^N I(\widehat{\theta}_{n,K})$, where each matrix $I(\widehat{\theta%
}_{n,K})$ is computed using the result of Proposition \ref%
{prop:prop2}, for $K=2000$ and $N=200$, compared to the lower bound $\sqrt{\Sigma(\theta^0)}$ (\ref{6a}) of the covariance matrix 
$\sqrt{I^{-1}(\widehat{\theta}_K)}$. The last column
lists the p-value of Kolmogorov-Smirnov statistic for goodness-of-fit
between empirical CDF of standardized biases and N(0,1) CDF. }
\label{table:eststdev}
\end{table}

\subsection{Simulation results}

We independently obtained $N=200$ sets of $K=800,1200$ and $2000$ sample
paths with each sample path generated in the way described in Section \ref%
{subsec:MCSim}. We chose $\phi ^{0}=(1/2,1/2,1/2)$ to be the initial value
of $(\phi _{1,1},\phi _{2,1},\phi _{3,1}),$ and chose the initial values of $%
Q_{1}$ and $Q_{2}$ as described in Section 3.1. The EM algorithm was run
until convergence criterion $||\theta ^{p+1}-\theta ^{p}||<10^{-4}$ was
achieved and was repeated $200$ times for each $K$ sample paths.

The simulation results were evaluated using Bias and the Root Mean Squared
Error (RMSE) and are reported in Table \ref{table:biasmse}. We observe that
as the sample size $K$ increases both Bias and RMSE decrease confirming
consistency of the EM estimates. We see from Table \ref{table:eststdev} that
the standard errors obtained from the simulation are close to the
theoretical standard errors i.e., the ones obtained from the inverse of
Fisher information. We next confirmed that for $K=2000$ and $N=200$, the
distribution of $\widehat{\theta }_{K}$ is approximately normal using the
Kolmogorov-Smirnov (KS) test as follows. For each element of $\theta ,$ the
simulation yields a random sample of $200$ biases, with each bias obtained
from a set of sample paths of size $2000$. We then standardized the biases
by dividing them by their standard error. The fit of the standardized biases
to the standard normal cfd was assessed using KS test. The p-value of the
KS\ test, reported for each element in Table \ref{table:eststdev}, confirms
the large sample normality of its estimator.


\section{Application to ventICU dataset}

\label{sec:appl}

\subsection{Data and the choice of the mixture model}

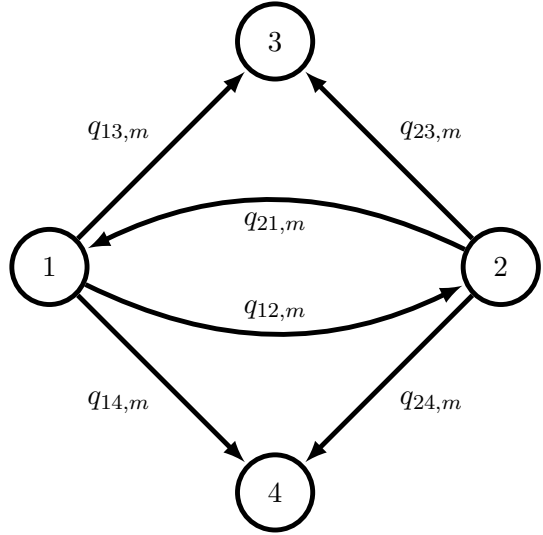
\begin{figure}[t]
\begin{center}
\begin{tikzpicture}[font=\sffamily]

        \tikzset{node style/.style={state,
                                    minimum width=1cm,
                                    line width=0.65mm,
                                    fill=white!20!white}}

          \tikzset{My Rectangle3/.style={rectangle, draw=brown, fill=white, thick,
    prefix after command= {\pgfextra{\tikzset{every label/.style={blue}}, label=below}}
    }
}

        \node[node style] at (3, 0)     (s1)     {$1$};
         \node[node style] at (6, 3)     (s3)     {$3$};
         \node[node style] at (6, -3)     (s4)     {$4$};
         \node[node style] at (9, 0)    (s2)     {$2$};

        \draw[every loop,
              auto=right,
              line width=0.65mm,
              >=latex,
              draw=orange,
              fill=orange]

             (s1)  edge[->, auto=left] node {$q_{13,m}$} (s3)
             (s1)  edge[->, auto=right] node {$q_{14,m}$} (s4)

              (s2)  edge[->, auto=right]                    node {$q_{23,m}$}  (s3)
              (s2)  edge[->, auto=left]                    node {$q_{24,m}$}  (s4)

             (s2)  edge[bend right=26, auto=left]                node {$q_{21,m}$}  (s1)
             (s1)  edge[bend right=26, auto=left]                node {$q_{12,m}$}  (s2);

 \end{tikzpicture}
\end{center}
\caption{State diagram for the m-th regime of ventICU mixture model.}
\label{fig:mixabs}
\end{figure}

This section applies the methods developed in the paper to the ventICU
dataset from Appendix D in Cook and Lawless (2018). The ventICU dataset
comes from a prospective cohort study of patients in an intensive care unit
(ICU), and contains information on the occurrence of infections and the need
for mechanical ventilation along with discharge and death times. Cook and
Lawless suggest that the multi-state model in Figure \ref{fig:mixabs} would
be suitable for examining the relation between mechanical ventilation status
and risk of death or discharge for the sample of 747 patients, but do not
provide any further analysis of the ventICU dataset. In Figure \ref{fig:mixabs},
state $1$ represents being off mechanical ventilator; state $2$ being on
mechanical ventilator; state $3$ discharge and state $4$ death. The numbers
of transitions between states are exhibited in Table 4 from which we see
that at the end of the study 733 patients were either discharged or died,
and 14 were still in the hospital. Thus, these 14 patients have right-censored times of stay in the hospital. The focus of our analysis is on identifying the subgroups of patients characterized by
different relations between mechanical ventilation status and risk of death
or discharge.

We use Akaike information criterion (AIC) to decide on the choice of the
g-mixture for the ventICU data. We estimate six g-mixtures for the ventICU
data with the number of regimes $M$ ranging from 1 (single Markov process)
to 6. Table 5 shows that $\mathrm{AIC}_{M}=2|\theta _{M}|-2$$\log L(\widehat{%
\theta }_{M})$, where $|\theta _{M}|$ is the number of parameters in the
M'th model, is the smallest for $M=2,$ and thus we use the 2-regime mixture
in the further analysis of the ventICU\ data.


\begin{table}[th]
\centering
\begin{tabular}{rrrrrr}
\hline
& 1 & 2 & 3 & 4 & \textrm{Censored} \\ \hline
1 & 0 & 75 & 585 & 21 & 5 \\ 
2 & 319 & 0 & 72 & 55 & 9 \\ \hline
\end{tabular}%
\caption{Observed transitions among different states and the numbers of
patients right-censored in states 1 and 2.}
\label{table:trans}
\end{table}

\begin{table}[ht!]
\centering 
\begin{tabular}{cccc}
\hline\hline
Model & AIC & $\log \mathcal{L}(\widehat{\theta})$   \\[0.5ex] \hline
Markov & 9648.185 & -4817.092   \\[1pt] 
2 Mixture & \textbf{9611.868} & -4790.934   \\[1pt] 
3 Mixture & 9611.991 & -4782.996   \\[1pt] 
4 Mixture & 9618.506 & -4778.253   \\[1pt] 
5 Mixture & 9628.735 & -4775.367   \\[1pt] 
6 Mixture & 9641.453 & -4773.727   \\[1ex] \hline
\end{tabular}
\caption{Summary of model statistics AIC and $\log \mathcal{L}(\widehat{\theta})$.}
\label{table:AICBIC2}
\end{table}

\subsection{Parameter estimates of the two-regime g-mixture model}

From the EM algorithm, the estimated numbers $\widehat{B}_{i,m}$ of patients
starting in the state $i$ and making transitions according to Markov process 
$X_{m},m=1,2,$ and the regime probabilities $\widehat{\phi }_{i,m}=\widehat{B%
}_{i,m}/B_{i}$ with their standard errors are reported in Table \ref%
{table:BimEst}. 
\begin{table}[th]
\centering 
\begin{tabular}{cccccc}
\hline\hline
State(i) & $B_i$ & $\widehat{B}_{i,1}$ & $\widehat{B}_{i,2}$ & $\widehat{\phi%
}_{i,1}\,(SE)$ & $\widehat{\phi}_{i,2}\,(SE)$ \\[0.5ex] \hline
1 & 367 & 230.75 & 136.25 & 0.6287 (.1674) & 0.3713 (.1674) \\[3pt] 
2 & 380 & 209.16 & 170.84 & 0.5504 (.1391) & 0.4496 (.1391) \\[1ex] \hline
\end{tabular}
\caption{Estimates of $B_{i,m}$'s and $\phi _{i,m}$'s with their
standard errors for the ventICU dataset.}
\label{table:BimEst}
\end{table}
We will refer to the patients estimated to evolve according to $X_{m}$ as $%
X_{m}$ patients. We see from Table \ref{table:BimEst} that initially there
were $367$ patients in state $1$ and $380$ in state $2$. Using this
information together with $\widehat{D}_{m}=$diag$(\widehat{\phi }_{1,m},%
\widehat{\phi }_{2,m})$, $m=1,2$, we obtain the estimated total number of $%
X_{1}$ patients, by summing the elements of the vector $\widehat{C}%
_{1}\equiv $($367,380)\widehat{D}_{1}=(230.75,209.16)$ to be approximately $%
440$. Similarly summing the elements of $\widehat{C}_{2}\equiv (367,380)%
\widehat{D}_{2}=(136.25,170.84),$ we get the estimated total number of $%
X_{2} $ patients to be approximately $307$. The estimates of the intensity
matrices $Q_{1}$ and $Q_{2}$ of the Markov processes $X_{1}$ and $X_{2}$
with the standard errors (in parentheses), computed using the results in
Section 4.2, are 
\begin{equation*}
\widehat{Q}_{1}=%
\begin{array}{c}
1 \\ 
\\ 
2 \\ 
\end{array}%
\left( 
\begin{array}{ccccc}
-0.16112 & 0.01657 & 0.14455 & 0.00000  \\ 
& (.00370) & (.01045) &    \\ 
0.12071 & -0.13832 & 0.01405 & 0.00356   \\ 
(.03143) &  & (.00558) & (.00532)  
\end{array}%
\right) ,
\end{equation*}%
and 
\begin{equation*}
\widehat{Q}_{2}=%
\begin{array}{c}
1 \\ 
\\ 
2 \\ 
\end{array}%
\left( 
\begin{array}{ccccc}
-0.11594 & 0.01441 & 0.09102 & 0.01051   \\ 
& (.00470) & (.01600) & (.00416)   \\ 
0.02309 & -0.04550 & 0.01094 & 0.01147   \\ 
(.00599) &  & (.00286) & (.00248)  
\end{array}%
\right) ,
\end{equation*}%
respectively, where we omitted the last two rows of zeros, which correspond
to the two absorbing states. To prevent having singularity in the inverse of 
$I(\widehat{\theta }),$\ we excluded $\widehat{q}_{14,1}$\ from the vector $%
\widehat{q}_{1}$\ in the estimation of the standard errors, and assumed that
the true value of $q_{14,1}=0.$ We see that the regime 2 death intensities
are much higher from both initial states than the regime 1 death intensites:
in fact regime 1 death intensity for patients who were initially not on
ventilator is zero. We also see that regime 1 discharge intensities are
higher from both initial states compared to the similar regime 2
intensities. In both regimes, not being initially on ventilator results in a
larger discharge intensity compared to the death intensity.

\subsubsection{Absorption probabilities}

To gain a better understanding of the differences between the two regimes,
we compare their absorption probabilities $\widehat{f}_{ij,m}$, $m=1,2$,
from state $i=1,2$ into states $j=3,4$. The standard equations for these
probabilities are derived based on the transition matrices of the discrete
time Markov chains embedded into the Markov processes $X_{1}$ and $X_{2},$
and are given in part C of Supplementary material. Here we just state the
results: the absorption probability matrices for the two regimes, denoted by 
$\widehat{F}_{1}$ and $\widehat{F}_{2},$ with entries $\widehat{f}_{ij,1}$
and $\widehat{f}_{ij,2}$, $(i=1,2),(j=3,4$), are 
\begin{equation*}
\widehat{F}_{1}=%
\begin{array}{c}
1 \\ 
2%
\end{array}%
\left( 
\begin{array}{ccc}
0.9971 & 0.0029   \\ 
0.9717 & 0.0283  
\end{array}%
\right) \;\;\text{and}\;\;\widehat{F}_{2}=%
\begin{array}{c}
1 \\ 
2%
\end{array}%
\left( 
\begin{array}{ccc}
0.8698 & 0.1302   \\ 
0.6818 & 0.3182  
\end{array}%
\right) ,
\end{equation*}%
respectively. Comparing $\widehat{F}_{1}$ with $\widehat{F}_{2}$, we see a
striking difference in the estimated probability of eventual death in the
two regimes. In regime 1, the probability of death for a patient initially
in state 1 (not on ventilator) is very small (0.0029), whereas this
probability\ for a patient initially in state 1 in regime 2 is large
(0.1302). For a patient initially in state 2 (on ventilator), the
probability of death is about 11 times larger in regime 2 compared to regime
1. As a result of very large differences in the death probabilities between
two regimes, we also observe large differences in their eventual discharge
probabilities.

To translate above results into{\ those involving patients' absorption
frequencies in each regime, we would want to pre-multiply }$\widehat{F}_{m}$
matrix by the row vector $\widehat{C}_{m}$ to obtain a row vector showing
the estimated number of $X_{m}$ patients who were eventually discharged, or
died. By doing so we would overestimate the absorption frequencies by 14
patients who were right-censored at the end of the study. Among those
patients, 5 were initially in state 1 and 9 in state 2. To compute
absorption frequencies, we have to subtract the 14 patients from the $%
\widehat{C}_{m}$ vectors, which we do as follows. According to $\widehat{D}%
_{m}$ matrices$,$ the 5 patients with initial state 1, contribute $5$($%
\widehat{\phi }_{11})=5(0.6287)$ patients to regime 1 patients and $5$($%
\widehat{\phi }_{12})=5(0.3713)$ to regime 2 patients. Similarly, the
patients initially in state 2, contribute $9(\widehat{\phi }_{21})=9(0.5504)$
patients to regime 1 patients and $9(\widehat{\phi }_{22})=9(0.4496)$ to
regime 2 patients. Hence, vector $\widehat{C}_{1}=(230.75,209.16)$ has to be
modified to become vector $\widehat{C}_{1,U}$ describing the regime 1
uncensored patients' absorption frequencies by initial state: 
\begin{equation*}
\widehat{C}_{1,U}=\widehat{C}_{1}-[5(0.6287),9(0.5504)]=(227.6065,204.2064),
\end{equation*}%
and vector $\widehat{C}_{2}=(136.25,170.84)$ has to be modified to become
vector $\widehat{C}_{2,U}$ describing the regime 2 uncensored patients'
absorption frequencies by initial state: 
\begin{equation*}
\widehat{C}_{2,U}=\widehat{C}_{2}-[5(0.3713),9(0.4496)]=(134.3935,166.7936)
\end{equation*}%
We can now compute the regime 1 absorption frequencies: 
\begin{equation*}
\widehat{C}_{1,U}\widehat{F}_{1}\approx
(226.95,0.66)+(198.43,5.78)=(425.38,6.44),
\end{equation*}%
where the first vector shows that absorption frequencies from state 1 and
the second those frequencies from state 2 among estimated $X_{1}$ patients.
Summing the two vectors tells us that in regime 1, about 425 patients were
eventually discharged and about 6 died. For regime 2, we have%
\begin{equation*}
\widehat{C}_{2,U}\widehat{F}_{2}\approx
(116.9,17.5)+(113.72,53.07)=(230.62,70.57),
\end{equation*}%
where the interpretation of the vectors is analogous to the one for the
regime 1. We see that the estimated total number of deaths is $77.01\approx
77$ which is one more than the observed number of deaths, see Table 5, and
the estimated total number of discharges is about $656,$ which is one less
than the observed number of discharges. This difference is likely due to the
rounding errors. The regime 1 death rate is $6.44/431.82\approx 0.015$ or$\
1.5\%$ whereas the regime 2 death rate is about ${70.57/301.2=0.234}${\ or }$%
{23.4\%}${.} Consequently, the rate of discharge from both initial states is
much larger in regime 1 compared to regime 2. In both regimes most of the
patients who died were initially on ventilator (state 2).\ The proportion of
deaths when initially on ventilator is $53.07/70.57$ or about $0.75$ in
regime 2 and $5.78/6.44\approx 0.9$ in regime 1.

Thus, the g-mixture model has identified two regimes corresponding to high
risk (regime 2) and low (regime 1) risk of death, with the\ patients in the
high risk regime having about 15 times higher death rate than those in the
low risk regime.

\subsection{The likelihood ratio test of the constrained vs unconstrained
mixture}

We want to see, if for ventICU data, there is a benefit in using the general
mixture proposed in this paper over the constrained mixture considered by
Frydman (2005). In the two-regime c-mixture, the two transition intensity
matrices $Q_{1}$ and $Q_{2}$ are constrained by assuming that $Q_{1}=\Gamma
Q_{2},$ where $\Gamma =$diag($\gamma _{1,1}$, $\gamma _{2,1}).$ This
constraint implies that the two regimes have the same embedded Markov chains
and thus also the same absorption probabilities. The test of c-mixture vs
g-mixture can be formulated as $H_{0}:$ $Q_{1}=\Gamma Q_{2}$ vs $H_{a}:Q_{1}$
and $Q_{2}\ $are unrestricted. To carry out the test, we use the likelihood
ratio statistic, -$2\log \Lambda :$ 
\begin{equation*}
\Lambda =\frac{\mathcal{L}_{\text{c-Mixture}}(\widehat{\theta }_{c})}{%
\mathcal{L}_{\text{g-Mixture}}(\widehat{\theta }_{g})},
\end{equation*}%
where $\widehat{\theta }_{c}$ and $\widehat{\theta }_{g}$ are the MLEs of
the vector parameters $\theta _{c}=$$\{\phi _{1,1,}^{c}\phi _{2,1}^{c}\}\cup
\{q_{ij}^{c},i\in E,j\in S,i\neq j\}\cup \{\gamma _{1,1},\gamma _{2,1}\}$
and $\theta _{g}=\{\phi _{1,1,}\phi _{2,1}\}\cup \{q_{ij,m},1\leq m\leq
2,i\in E,j\in S,i\neq j\},$ respectively, where $E=\{1,2\}$ and $\Delta
=\{3,4\}.$ We see that a general mixture has 14 parameters while the
constraint one has 10, which means that, under $H_{0},$ $-2\log \Lambda $
has $\chi ^{2}$ distribution with 4 degrees of freedom. To evaluate $-2\log
\Lambda ,$ we note that the likelihood function of a general mixture, $L_{%
\text{g-Mixture}}(\theta _{g}),$ is given in (\ref{a2}) and the likelihood
function of the constrained mixture is of the form 
\begin{eqnarray*}
\mathcal{L}_{\text{c-Mixture}}(\theta _{c}) &=&\left(
\prod_{k=1}^{K}\prod_{i\in E}\pi _{i}^{B_{i}^{k}}\right)
\prod_{k=1}^{K}\sum_{m=1}^{M}\Bigg[\prod_{i\in E}\left( \phi
_{i,m}\right) ^{B_{i}^{k}} \\
&&\times \prod_{i\in E}\left\{ \prod_{j\neq i,j\in S}(\gamma
_{i,m}q_{ij})^{N_{ij}^{k}}\,\exp \Big(-\sum_{j\neq i,j\in S}\gamma
_{i,m}q_{ij}T_{i}^{k}\Big)\right\} \Bigg],
\end{eqnarray*}%
where the factor in parentheses involving $\pi _{i}^{\prime }s$ is the same
as the factor involving $\pi _{i}^{\prime }s$ in $L_{\text{g-Mixture}%
}(\theta _{c})$ and thus $\pi _{i}^{\prime }s$ play no role in the
evaluation of $-2\log \Lambda .$ From Table \ref{table:AICBIC2}, $\log (L_{%
\text{g-Mixture}}(\widehat{\theta }_{g}))=-4790.934$, and from the EM
algorithm applied to fit the c-mixture to ventICU data, $\log (L_{\text{%
c-Mixture}}(\widehat{\theta }_{c}))=-4797.926$. Thus, $-2\log \Lambda
=13.984 $ with the p-value of 0.00735 show that we can reject c-mixture in
favor of g-mixture at $\alpha =0.01.$

\section{Concluding remarks}

We proposed and estimated a new unconstrained mixture of Markov processes.
We showed the consistency and asymptotic normality of the estimators of the
mixture's parameters and obtained the finite sample standard errors of the
estimates. The simulation study verified that the estimation was accurate
and confirmed the asymptotic properties of the estimators. The application
of the proposed mixture to VenICU illustrated its usefulness in identifying
subpopulations and its dominance over the constrained mixture. We believe
that the unconstrained mixture will dominate the constrained one in many
other data sets arising from heterogeneous populations. We intend to extend
the proposed general mixture in a number of ways which include incorporation
of covariates into our continuous observation time framework, and developing
estimation from observing the mixture at discrete time points in the
presence of covariates. The estimation of a discretely observed Markov jump
process without covariates has been considered by Bladt and S{\o }rensen
(2005, 2009), Inamura (2006), Mostel et al. (2020), and Pfeuffer et. al.
(2019), among others. The mixture proposed here and the mixtures of
finite-state continuous-time Markov processes arising from the above
potential extensions should be useful in a variety of contexts in which
modeling of the population heterogeneity is important.

\section*{Acknowledgements}

Part of this work was carried out while Budhi Surya was visiting Department
of Technology, Operations and Statistics at New York University Stern School
of Business in September 2019. He acknowledges Faculty Strategic Research
Grant No. 20859 of Victoria University of Wellington and hospitality
provided by the NYU Stern.

\nocite{*}
\bibliographystyle{acm}

\begin{thebibliography}{99}
\bibitem{Akaike} Akaike, H. (1973) Information theory and extension of the
maximum likelihood principle. In \textit{Proceedings of the 2nd
International Symposium on Information Theory}, 267-281, eds. B.N Petrov and
F Csaki, Akademia Kiado, Budapest.

\bibitem{Albert} Albert, A. (1962) Estimating the infinitesimal generator of
a continuous time, finite state Markov process. \textit{Annals of
Mathematical Statistics} \textbf{38}, 727-753.



\bibitem{Bladt2009} Bladt, M. \& S{\o }rensen, M. (2009) Efficient
estimation of transition rates between credit ratings from observations at
discrete time points. \textit{Quantitative Finance} \textbf{9} (2), 147-160.

\bibitem{Bladt2005} Bladt, M. \& S{\o }rensen, M. (2005) Statistical
inference for discretely observed Markov jump processes. \textit{Journal of
the Royal Statistical Society Series B} \textbf{67} (3), 395-410.

\bibitem{Blumen} Blumen, I., Kogan, M. \& McCarthy, P.J. (1955) The
industrial mobility of labor as a probability process. \textit{Cornell
Studies in Industrial and Labor Relations}, Vol. 6, Ithaca, N.Y., Cornell
University Press.


\bibitem{Cipollini} Cipollini, F., Ferretti, C. \& Ganugi, P. (2012) Firm
size dynamics in an industrial district: The mover-stayer model in action. 
\textit{Advanced Statistical Methods for the Analysis of Large Data-Sets,}
ed. A. Di Ciaccio, M. Coli, and J. M. Angulo Ibanez,443--452. Berlin:
Springer.

\bibitem{Cook} Cook, R.J. \& Lawless, J.F. (2018) \textit{Multistate
Models for the Analysis of Life History Data}. CRC Press, Taylor \& Francis.

\bibitem{Cook2002} Cook, R.J., Kalbfleisch, J.D., \& Yi, G.Y. (2002) A
generalized mover--stayer model for panel data. \textit{Biostatistics }%
\textbf{3} (3), 407-420.


\bibitem{Dempster} Dempster, A.P., Laird, N.M., \& Rubin, D.B. (1977)
Maximum likelihood from incomplete data via the EM algorithm (with
discussion). \textit{Journal of the Royal Statistical Society Series B} 
\textbf{39}, 1-38.

\bibitem{Feng} Feng, C., Wang, H., Han, Y., Xia, Y., \& Tu, X.M. (2013)
The mean value theorem and Taylor's expansion in statistics. \textit{The
American Statistician} \textbf{67}(4), p. 245-248.

\bibitem{Ferguson} Ferguson, T.S. (1996) \textit{A Course in Large Sample
Theory}. Chapman \& Hall.

\bibitem{Ferretti} Ferretti, C., Gabbi, G., Ganugi, P., Sist, F. \& Vozzella, P. (2019) Credit Risk Migration and Economic Cycles. \textit{Risks} \textbf{7} (4).

\bibitem{Fougere} Foug\'ere, D. \& Kamionka, T. (2003) Bayesian inference
for the mover-stayer model in continuous time with an application to labour
market transition data. \textit{Journal of Applied Econometrics} \textbf{18}
(6), 697-723.

\bibitem{Frydman2019} Frydman, H., Matuszyk, A., Li, C., \& Zhu, W. (2019)
Mover-stayer model with covariate effects on stayer's probability and
mover's transitions. \textit{Applied Stochastic Models in Business and
Industry} \textbf{35} (5), 1171-1184.

\bibitem{Frydman2018} Frydman, H. \& Matuszyk, A. (2018). Estimation and
status prediction in a discrete mover-stayer model with covariate effects on
stayer's probability. \textit{Applied Stochastic Models in Business and
Industry} \textbf{34} (2), 196-205.

\bibitem{Frydman2008} Frydman, H. \& Schuermann, T. (2008) Credit rating
dynamics and Markov mixture models. \textit{Journal of Banking and Finance} 
\textbf{32}, 1062-1075.

\bibitem{Frydman2005} Frydman, H. (2005). Estimation in the mixture of
Markov chains moving with different speeds. \textit{Journal of the American
Statistical Association} \textbf{100}, 1046-1053.

\bibitem{Frydman2004} Frydman, H. \& Kadam, A. (2004) Estimation in the
continuous time mover-stayer model with an application to bond ratings
migration. \textit{Applied Stochastic Models in Business and Industry} 
\textbf{20} (2), 155-170.

\bibitem{Frydman} Frydman, H. (1984) Maximum likelihood estimation in the
mover stayer model. \textit{Journal of the American Statistical Association} 
\textbf{79} (387), 632-638.


\bibitem{Horn} Horn, R.A. \& Johnson, C.R. (2013) \textit{Matrix Analysis}. Cambridge University Press.

\bibitem{Inamura} Inamura, Y. (2006) Estimating continuous time transition
matrices from discretely observed data. \textit{Bank of Japan Working Paper
Series}. \url{www.boj.or.jp/en/research/wps_rev/wps_2006/data/wp06e07.pdf}

\bibitem{Jiang} Jiang, S. \& Cook, R.J. (2019) Score tests based on a
finite mixture model of Markov processes under intermittent observation. 
\textit{Statistics in Medicine} \textbf{38} (16), 3013-3025.

\bibitem{Louis1982} Louis, T. A. (1982) Finding the observed information
matrix when using the EM algorithm. \textit{Journal of the Royal Statistical
Society Series B} \textbf{44}, 226-233.

\bibitem{Magnus} Magnus, J.R. \& Neudecker, H. (2007) \textit{Matrix
Differential Calculus with Applications in Statistics and Econometrics}, 3rd
Ed, John Willey \& Sons.

\bibitem{Mostel} Mostel, L., Pfeuffer, M., \& Fischer, M. (2020) Statistical
inference for Markov chains with applications to credit risk. \textit{Computational Statistics} \textbf{35}, 1659-1684.


\bibitem{Fruhwirth} Pamminger, C. \& Fruhwirth-Schnatter, S. (2010)
Model-based clustering of categorical time series. \textit{Bayesian Analysis}
\textbf{5} (2), 345-368.

\bibitem{Pfeuffer2019} Pfeuffer, M., Mostel, L., \& Fischer, M. (2019) An
extended likelihood framework for modelling discretely observed credit
rating transitions. \textit{Quantitative Finance} \textbf{19} (1), 93-104.

\bibitem{Saint-Cyr} Saint-Cyr, L. D. F., \& Piet, L. (2017) Movers and
stayers in the farming sector: Accounting for unobserved heterogeneity in
structural change. \textit{Journal of the Royal Statistical Society Series C}
\textbf{66}, 777--795.


\bibitem{Schervish} Schervish, M.J. (1995) \textit{Theory of Statistics}.
Springer.

\bibitem{Shen} Shen, H. \& Cook, R.J. (2014) A dynamic Mover--Stayer model
for recurrent event processes subject to resolution.\textit{\ Life time data
analysis }\textbf{20, }404-423.

\bibitem{Sigman} Sigman K. (2007) Simulation of Markov Chains. \textit{%
Lecture Notes}. \url{www.columbia.edu/~ks20/4703-Sigman/4703-07-Notes-MC.pdf}

\bibitem{Surya2018} Surya, B.A. (2018) Distributional properties of the
mixture of continuous-time absorbing Markov chains moving at different
speeds. \textit{Stochastic Systems} \textbf{8}, 29-44.

\bibitem{Tabar} Tabar, L. Fagerberg, G. \& Chen, H. (1996) Tumor
development, histology and grade of breast cancers: Prognosis and
progression. \textit{Journal of Cancer} \textbf{66}, 413-419.



\bibitem{Wu} Wu, C.F.J. (1983) On the convergence properties of the EM
algorithm. \textit{The Annals of Statistics} \textbf{11} (1), 95-103.

\bibitem{Yi} Yi, G.Y., He, W. \& He, F. (2017) Analysis of panel data under
hidden mover--stayer models. \textit{Statistics in Medicine} \textbf{36}
(20), 3231-3243.
\end{thebibliography}

\appendix

\end{document}


\title{Supplementary material for : \\ Statistical inference for a mixture of Markov jump processes}
\author{Halina Frydman \thanks{%
Department of Technology, Operations and Statistics, Stern School of
Business, New York University, New York, NY 10012, USA, email:
hfrydman@stern.nyu.edu}
\, Budhi Surya \thanks{%
School of Mathematics and Statistics, Victoria University of Wellington,
Gate 6, Kelburn PDE, Wellington 6140, New Zealand, email:
budhi.surya@vuw.ac.nz} \thanks{%
The author acknowledges financial support through PBRF research grant No.
220859.} }

\date{10 April 2022}

\maketitle

\appendix

\section{Proof of Theorem 2 }

\subsection{Unconditional moments of $B_{i,m}$, $N_{ij,m}$ and $T_{i,m}$}

For the assertion of the theorem, we need the unconditional moments of $B_{i,m}$, $N_{ij,m}$ and $T_{i,m}$.
\begin{lemma}\label{lem:lemB2}
For $i\in E$, $j\in S$, and $1\leq m\leq M$, we have for a fixed $T>0$,
\begin{eqnarray*}
\mathbb{E}_{\theta}\big[\Phi_{k,m}N_{ij}^k\big]&=&q_{ij,m}\int_0^T \pi^{\top}D_me^{Q_m u} e_i du,\\
\mathbb{E}_{\theta}\big[\Phi_{k,m}T_i^k\big]&=&\int_0^T \pi^{\top}D_me^{Q_m u} e_i du,
\end{eqnarray*}
while for $T$ being the absorption time of $X$, 
\begin{eqnarray*}
\mathbb{E}_{\theta}\big[\Phi_{k,m}N_{ij}^k\big]&=&q_{ij,m}\int_0^{\infty} \pi^{\top}D_me^{Q_m u} e_i du,\\
\mathbb{E}_{\theta}\big[\Phi_{k,m}T_i^k\big]&=&\int_0^{\infty} \pi^{\top}D_me^{Q_m u} e_i du.
\end{eqnarray*}
\end{lemma}

\subsection*{For a fixed time $T>0$}

The proofs are based on adapting similar arguments to the proof of
Theorem 5.1 in Albert (1962) by dividing the interval $[0,T]$
into $n$ equal parts of length $h=T/n$. Then, by dominated convergence
theorem,
\begin{eqnarray*}
\mathbb{E}_{\theta}\big[\Phi_{k,m} N_{ij}^k\big]&=&\mathbb{E}_{\theta}\Big[\sum_{\ell=0}^{\infty}\mathbf{1}_{\{\Phi_{k,m}=1,X_{\ell h}^k=i, X_{(\ell +1)h}^k=j,\,(\ell+1)h\leq T\}}\Big]\\
&&\hspace{-2cm}= \sum_{\ell=0}^{n-1} \sum_{v\in \mathbb{S}} \mathbb{P}_{\theta}\big\{X_{\ell h}^k=i, X_{(\ell +1)h}^k=j, \Phi_{k,m}=1, X_0^k=v\big\}\\
&&\hspace{-2cm}= \sum_{\ell=0}^{n-1} \sum_{v\in \mathbb{S}} \mathbb{P}_{\theta}\{X_0^k=v\}\mathbb{P}_{\theta}\{\Phi_{k,m}=1\big\vert X_0^k=v\}\\
&&\hspace{-0.5cm}\times\mathbb{P}_{\theta}\{X_{\ell h}^k=i\big\vert \Phi_{k,m}=1,X_0^k=v\}\\
&&\hspace{0.5cm}\times
\mathbb{P}_{\theta}\{X_{(\ell+1)h}^k=j\big\vert \Phi_{k,m}=1,X_{\ell h}^k=i, X_0^k=v\}\\
&&\hspace{-2cm}=\sum_{\ell=0}^{n-1} \sum_{v\in \mathbb{S}} \pi_v \phi_{v,m} e_v^{\top} e^{Q_m \ell h} e_i q_{ij,m} h\\
&&\hspace{-2cm}=\sum_{\ell=0}^{n-1} \pi^{\top} D_m e^{Q_m \ell h} e_i q_{ij,m} h  \stackrel{h\rightarrow 0}{\longrightarrow} q_{ij,m}\int_0^T \pi^{\top} D_m e^{Q_m u} e_i du.
\end{eqnarray*}
For the second statement, let $Y_i^k(u)=\mathbf{1}_{\{X_u^k=i\}}$. Then by Fubini's theorem,
\begin{eqnarray*}
\mathbb{E}_{\theta}\big[\Phi_{k,m}T_i^k\big]&=&\mathbb{E}_{\theta}\Big[\int_0^T \Phi_{k,m} Y_i^k(u) du\Big]\\
&=&\int_0^T \mathbb{P}_{\theta}\big\{ X_u^k=i, \Phi_{k,m}=1\big\}du\\
&=&\int_0^T \sum_{v\in\mathbb{S}}  \mathbb{P}_{\theta}\big\{ X_u^k=i, \Phi_{k,m}=1, X_0^k=v\big\}du\\
&=&\int_0^T \sum_{v\in\mathbb{S}} \mathbb{P}_{\theta}\{X_0^k=v\}\mathbb{P}_{\theta}\{\Phi_{k,m}=1\vert X_0^k=v\}\\
&&\hspace{0.5cm}\times \mathbb{P}_{\theta}\{X_u^k=i\big\vert \Phi_{k,m}=1,X_0^k=v\}\\
&=&\int_0^T \sum_{v\in\mathbb{S}} \pi_v\phi_{v,m} e_v^{\top} e^{Q_m u} e_i du\\
&=& \int_0^T \pi^{\top} D_m e^{Q_m u} e_i du,
\end{eqnarray*}
which completes the proof of the claim for fixed $T>0$. $\blacksquare$

\subsection*{For an absorption time $T$}

Since $T$ is the absorption time of $X$, there are no $i\rightarrow j$ transitions from state $i\in E$ to $j\in S$ after $T$. Therefore, for a given $h>0$, we have
\begin{eqnarray*}
\mathbb{E}_{\theta}\big[\Phi_{k,m} N_{ij}^k\big]&=&\mathbb{E}_{\theta}\Big[\sum_{\ell=0}^{\lfloor T/h\rfloor -1} \Phi_{k,m} X_{\ell}^k(i,j)\Big]  \nonumber\\
&=&\mathbb{E}_{\theta}\Big[\sum_{\ell=0}^{\lfloor T/h\rfloor -1} \Phi_{k,m} X_{\ell}^k(i,j)\Big] + 
\mathbb{E}_{\theta}\Big[\sum_{\ell=\lfloor T/h\rfloor}^{\infty} \Phi_{k,m} X_{\ell}^k(i,j)\Big] \nonumber\\
&=&\mathbb{E}_{\theta}\Big[\sum_{\ell=0}^{\infty} \Phi_{k,m} X_{\ell}^k(i,j)\Big] \nonumber\\
&=& \sum_{\ell=0}^{\infty}\mathbb{P}_{\theta}\{\Phi_{k,m}=1, X_{\ell h}^k=i, X_{(\ell+1)h}^k=j\} \\
&=& \sum_{\ell=0}^{\infty} \sum_{v\in S}\mathbb{P}_{\theta}\{X_0^k=v,\Phi_{k,m}=1, X_{\ell h}^k=i, X_{(\ell+1)h}^k=j\} \\
&=&  \sum_{\ell=0}^{\infty} \sum_{v\in S} \mathbb{P}_{\theta}\{X_0^k=v\} \mathbb{P}_{\theta}\{\Phi_{k,m}=1\big\vert X_0^k=v\}\\
&&\hspace{1cm}\times \mathbb{P}_{\theta}\{X_{\ell h}^k=i\big\vert \Phi_{k,m}=1, X_0^k=v\}\\
&&\hspace{1cm}\times \mathbb{P}_{\theta}\{X_{(\ell+1)h}^k=j\big\vert \Phi_{k,m}=1, X_{\ell h}^k=i, X_0^k=v\}\\
&=&  \sum_{\ell=0}^{\infty} \sum_{v\in S} \pi_v \phi_{v,m} e_v^{\top} e^{Q_m \ell h} e_i q_{ij,m} h\\
&=& q_{ij,m}\sum_{\ell=0}^{\infty} \pi^{\top} D_m e^{Q_m \ell h} e_i h \stackrel{h\rightarrow 0}{\longrightarrow} q_{ij,m}\int_0^{\infty} \pi^{\top} D_m e^{Q_m u} e_idu,
\end{eqnarray*}
which establishes the assertion of the first statement. 

\medskip

To prove the second statement, recall that since $T$ is the absorption time, there is zero total occupation time of $X$ in state $i\in E$ after $T$. Thus,
\begin{eqnarray*}
\mathbb{E}_{\theta}\big[\Phi_{k,m} T_i^k\big]&=& \mathbb{E}_{\theta}\Big[\int_0^T \mathbf{1}_{\{\Phi_{k,m}=1,X_u^k =i\}} du\Big]\\
&=& \mathbb{E}_{\theta}\Big[\int_0^T \mathbf{1}_{\{\Phi_{k,m}=1,X_u^k =i\}} du + \int_T^{\infty}  \mathbf{1}_{\{\Phi_{k,m}=1,X_u^k =i\}} du\Big]\\
&=& \mathbb{E}_{\theta}\Big[\int_0^{\infty} \mathbf{1}_{\{\Phi_{k,m}=1,X_u^k =i\}} du\Big]\\
&=& \int_0^{\infty} \mathbb{P}_{\theta}\big\{\Phi_{k,m}=1, X_u^k=i\big\}du\\
&=& \int_0^{\infty}\sum_{v\in S} \mathbb{P}_{\theta}\big\{X_0^k=v, \Phi_{k,m}=1, X_u^k=i\big\}du\\
&=& \int_0^{\infty}\sum_{v\in S} \mathbb{P}_{\theta}\{X_0^k=v\}\mathbb{P}_{\theta}\{\Phi_{k,m}=1\big\vert X_0^k=v\} \\
&&\hspace{2cm}\times \mathbb{P}_{\theta}\{X_u^k=i\big\vert \Phi_{k,m}=1, X_0^k=v\}du\\
&=&  \int_0^{\infty} \pi^{\top} D_m e^{Q_m u} e_i du,
\end{eqnarray*}
which establishes the second claim for absorption time. $\blacksquare$

\subsection{Unconditional second moments of $N_{ij,m}$, $T_{i,m}$ and $B_{i,m}$}\label{sec:appx}

This part will have a proposition with its proof. It generalizes the results of Theorem 5.1 of Albert (1962) to a mixture of Markov jump processes.  

\begin{proposition}\label{prop:B1}
Let $D_m$ be a $(w\times w)$diagonal matrix with diagonal
elements $\phi _{i,m}$, $i\in E$. The unconditional moments of $B_{i}^{k}$, $%
N_{ij}^{k}$, and $T_{i}^{k}$ are 
\begin{align}
\mathbb{E}_{\theta}\Big[\Phi _{k,m}B_{i^{\prime }}^{k}N_{ij}^{k}\Big] =&q_{ij,m}\pi
_{i^{\prime }}\phi _{i^{\prime },m}e_{i^{\prime }}^{\top
}\int_{0}^{T}e^{Q_{m}u}e_{i}du, \label{eq:1a} \tag{a}\\
\mathbb{E}_{\theta}\Big[\Phi _{k,m}B_{i^{\prime }}^{k}T_{i}^{k}\Big] =&\pi
_{i^{\prime }}\phi _{i^{\prime },m}e_{i^{\prime }}^{\top
}\int_{0}^{T}e^{Q_{m}u}e_{i}du, \label{eq:1b} \tag{b}\\
\mathbb{E}_{\theta}\Big[\Phi _{k,m}N_{ij}^{k}N_{i^{\prime }j^{\prime }}^{k}\Big]
=&q_{ij,m}\delta _{i}(i^{\prime })\delta _{j}(j^{\prime })\pi ^{\top }D_m%
\Big(\int_{0}^{T}e^{Q_{m}u}du\Big)e_{i}\nonumber\\
&\hspace{-1cm}+q_{ij,m}q_{i^{\prime }j^{\prime },m}\pi ^{\top }D_m\Bigg[%
\int_{0}^{T}\int_{0}^{u}e^{Q_{m}t}e_{i^{\prime }}e_{j^{\prime }}^{\top
}e^{Q_{m}(u-t)}e_{i}dtdu  \label{eq:1c} \tag{c}\\
&\hspace{2cm}+\int_{0}^{T}\int_{0}^{u}e^{Q_{m}t}e_{i}e_{j}^{\top
}e^{Q_{m}(u-t)}e_{i^{\prime }}dtdu\Bigg],  \nonumber \\
\mathbb{E}_{\theta}\Big[\Phi _{k,m}T_{i}^{k}T_{i^{\prime }}^{k}\Big] =&\pi ^{\top
}D_m\Bigg[\int_{0}^{T}\int_{0}^{u}e^{Q_{m}t}e_{i}e_{i}^{\top
}e^{Q_{m}(u-t)}e_{i^{\prime }}dtdu \label{eq:1d} \tag{d} \\
&\hspace{2cm}+\int_{0}^{T}\int_{0}^{u}e^{Q_{m}t}e_{i^{\prime }}e_{i^{\prime
}}^{\top }e^{Q_{m}(u-t)}e_{i}dtdu\Bigg], \nonumber \\
\mathbb{E}_{\theta}\Big[\Phi _{k,m}N_{ij}^{k}T_{i^{\prime }}^{k}\Big] =&q_{ij,m}\pi
^{\top }D_m\Bigg[\int_{0}^{T}\int_{0}^{u}e^{Q_{m}t}e_{i^{\prime
}}e_{i^{\prime }}^{\top }e^{Q_{m}(u-t)}e_{i}dtdu  \label{eq:1e} \tag{e}\\
&\hspace{2cm}+\int_{0}^{T}\int_{0}^{u}e^{Q_{m}t}e_{i}e_{j}^{\top
}e^{Q_{m}(u-t)}e_{i^{\prime }}dtdu\Bigg], \nonumber
\end{align}%
where we have denoted by $e_{i}=(0,\cdots ,1,\cdots ,0)$ a $(w\times 1)$unit
vector with value one on the $i$th element and zero otherwise.
\end{proposition}

\subsubsection*{Proof of (\ref{eq:1a})}

The proofs are based on adapting similar arguments to that of
Theorem 5.1 in Albert (1962) by dividing the interval $[0,T]$
into $n$ equal parts of length $h=T/n$. Then, by dominated convergence
theorem, $N_{ij}^{k}=\sum_{\ell =0}^{n-1}X_{\ell }^{k}(i,j)$ with $X_{\ell
}^{k}(i,j)=I(X_{\ell h}^{k}=i,X_{(\ell +1)h}^{k}=j),$ $(i,j)\in S$. Thus,  
\begin{eqnarray*}
\mathbb{E}_{\theta}\Big[\Phi _{k,m}B_{i^{\prime }}^{k}N_{ij}^{k}\Big] &=&\mathbb{E}_{\theta}\Big[\sum_{\ell =0}^{n-1}\Phi _{k,m}B_{i^{\prime }}^{k}X_{\ell }^{k}(i,j)\Big] \\
&&\hspace{-1.5cm}=\sum_{\ell =0}^{n-1}\mathbb{P}_{\theta}\Big\{X_{\ell
h}^{k}=i,X_{(\ell +1)h}^{k}=j,X_{0}^{k}=i^{\prime },\Phi _{k,m}=1\Big\} \\
&&\hspace{-1.5cm}=\sum_{\ell =0}^{n-1}\mathbb{P}_{\theta}\big\{X_{0}^{k}=i^{\prime }%
\big\}\mathbb{P}_{\theta}\big\{\Phi _{k,m}=1\big\vert X_{0}^{k}=i^{\prime }\big\}\mathbb{P}_{\theta}\big\{X_{\ell h}^{k}=i\big\vert\Phi _{k,m}=1,X_{0}^{k}=i^{\prime }\big\} \\
&&\hspace{0cm}\times \mathbb{P}_{\theta}\Big\{X_{(\ell +1)h}^{k}=j\Big\vert\Phi
_{k,m}=1,X_{\ell h}^{k}=i,X_{0}^{k}=i^{\prime }\Big\} \\
&&\hspace{-1.5cm}=\sum_{\ell =0}^{n-1}\pi _{i^{\prime }}\phi _{i^{\prime
},n}e_{i^{\prime }}^{\top }e^{Q_{n}\ell h}e_{i}q_{ij,n}h,
\end{eqnarray*}%
which converges by dominated convergence to the one claimed. $\blacksquare$

\subsubsection*{Proof of (\ref{eq:1b})}

By Fubini's theorem, Bayes' formula, and Markov property of $X_m$,
\begin{eqnarray*}
\mathbb{E}_{\theta}\big[\Phi_{k,m} B_{i^{\prime}}^k T_i^k\big] &=&\mathbb{E}_{\theta}\Big[\Phi_{k,m} B_{i^{\prime}}^k \int_0^T \mathbf{1}_{\{X_u^k=i\}}du\Big]\\
&&\hspace{-2cm}=\int_0^T \mathbb{P}_{\theta}\{ X_u^k=i, X_0^k=i^{\prime}, \Phi_{k,m}=1\}du \\
&&\hspace{-2cm}= \int_0^T \mathbb{P}_{\theta}\{X_0^k=i^{\prime}\}\mathbb{P}_{\theta}\{\Phi_{k,m}=1\big\vert X_0^k=i^{\prime}\}\mathbb{P}_{\theta}\{X_u^k=i \big\vert \Phi_{k,m}=1,X_0^k=i^{\prime}\}du,
\end{eqnarray*}
which indeed gives the second statement. $\blacksquare$

\subsubsection*{Proof of (\ref{eq:1c})}

Using representation $N_{ij}^k=\sum_{\ell=0}^{n-1} X_{\ell}^k(i,j)$, one can write 
\begin{eqnarray*}
\Phi_{k,m}N_{ij}^k N_{i^{\prime}j^{\prime}}^k&=&\sum_{\ell=0}^{n-1} \Phi_{k,m}X_{\ell}^k(i,j) X_{\ell}^k(i^{\prime},j^{\prime}) \\
&&+ \sum_{\ell=1}^{n-1} \sum_{r < \ell}\Phi_{k,m}X_{\ell}^k(i,j) X_{r}^k(i^{\prime},j^{\prime}) +  \sum_{\ell=0}^{n-2} \sum_{r> \ell} \Phi_{k,m}X_{\ell}^k(i,j) X_{r}^k(i^{\prime},j^{\prime})\\
&=& \sum_{\ell=0}^{n-1} \Phi_{k,m}X_{\ell}^k(i,j) X_{\ell}^k(i^{\prime},j^{\prime})\\
&&+ \sum_{\ell=1}^{n-1} \sum_{r=0}^{\ell -1} \Phi_{k,m}X_{\ell}^k(i,j) X_{r}^k(i^{\prime},j^{\prime})
+ \sum_{r=1}^{n-1} \sum_{\ell=0}^{r -1} \Phi_{k,m}X_{\ell}^k(i,j) X_{r}^k(i^{\prime},j^{\prime}).
\end{eqnarray*}
Therefore,
\begin{eqnarray}
\mathbb{E}_{\theta}\big[\Phi_{k,m} N_{ij}^k N_{i^{\prime} j^{\prime}}^k\big] &=&\mathbb{E}_{\theta}\Big[ \sum_{\ell=0}^{n-1} \Phi_{k,m}X_{\ell}^k(i,j) X_{\ell}^k(i^{\prime},j^{\prime})\Big] \nonumber\\
&&\hspace{-4cm}+\mathbb{E}_{\theta}\Big[ \sum_{\ell=1}^{n-1} \sum_{r=0}^{\ell -1} \Phi_{k,m}X_{\ell}^k(i,j) X_{r}^k(i^{\prime},j^{\prime})\Big]
+\mathbb{E}_{\theta}\Big[ \sum_{r=1}^{n-1} \sum_{\ell=0}^{r -1} \Phi_{k,m}X_{\ell}^k(i,j) X_{r}^k(i^{\prime},j^{\prime})\Big]. \label{eq:sum1}
\end{eqnarray}
Since an initial state is chosen randomly using probability $\pi$, the first expectation is obtained using Bayes' formula and the law of total probability,
\begin{eqnarray*}
\mathbb{E}_{\theta}\Big[ \sum_{\ell=0}^{n-1} \Phi_{k,m}X_{\ell}^k(i,j) X_{\ell}^k(i^{\prime},j^{\prime})\Big]&=&\sum_{\ell=0}^{n-1}\mathbb{E}_{\theta}\Big[ \Phi_{k,m} X_{\ell}^k (i,j) X_{\ell}^k (i^{\prime},j^{\prime})\Big]\\
&&\hspace{-4cm}=
\sum_{\ell=0}^{n-1} \mathbb{P}_{\theta}\big\{X_{\ell h}^k=i, X_{(\ell +1)h}^k=j, X_{\ell h}^k=i^{\prime}, X_{(\ell +1)h}^k=j^{\prime},\Phi_{k,m}=1\big\} \\
&&\hspace{-4cm}= \delta_i(i^{\prime})\delta_j(j^{\prime})
\sum_{\ell=0}^{n-1} \mathbb{P}_{\theta}\big\{ X_{\ell h}^k=i^{\prime}, X_{(\ell +1)h}^k=j^{\prime},\Phi_{k,m}=1\big\} \\
&&\hspace{-4cm}= \delta_i(i^{\prime})\delta_j(j^{\prime})
\sum_{\ell=0}^{n-1}\sum_{v\in S} \mathbb{P}_{\theta}\big\{ X_{\ell h}^k=i^{\prime}, X_{(\ell +1)h}^k=j^{\prime},\Phi_{k,m}=1, X_0^k=v\big\}\\
&&\hspace{-4cm}= \delta_i(i^{\prime})\delta_j(j^{\prime})
\sum_{\ell=0}^{n-1}\sum_{v\in S} \mathbb{P}_{\theta}\{X_0^k=v\}\mathbb{P}_{\theta}\{\Phi_{k,m}=1\big\vert X_0^k=v\}\\&&\hspace{-1cm}\times \mathbb{P}_{\theta}\{X_{\ell h}^k =i^{\prime}\big\vert \Phi_{k,m}=1, X_0^k=v\}\\&&\hspace{-1cm}\times \mathbb{P}_{\theta}\{X_{(\ell +1)h}^k=j^{\prime}\big\vert \Phi_{k,m}=1, X_{\ell h}^k=i^{\prime}, X_0^k=v\}\\
&&\hspace{-4cm}= \delta_i(i^{\prime})\delta_j(j^{\prime})
\sum_{\ell=0}^{n-1}\sum_{v\in S} \pi_v \phi_{v,m} e_v^{\top} e^{Q_m \ell h} e_{i^{\prime}} q_{i^{\prime} j^{\prime},m} h\\
&&\hspace{-4cm}= q_{i^{\prime} j^{\prime},m} \delta_i(i^{\prime})\delta_j(j^{\prime})
\sum_{\ell=0}^{n-1} \pi^{\top} D_m e^{Q_m \ell h} e_{i^{\prime}} h,
\end{eqnarray*}
which by dominated convergence theorem the last sum converges: 
\begin{eqnarray*}
q_{i^{\prime} j^{\prime},m} \delta_i(i^{\prime})\delta_j(j^{\prime})
\sum_{\ell=0}^{n-1} \pi^{\top} D_m e^{Q_m \ell h} e_{i^{\prime}} h \stackrel{h \rightarrow 0}{\longrightarrow} 
q_{i^{\prime} j^{\prime},m} \delta_i(i^{\prime})\delta_j(j^{\prime})  \int_0^T  \pi^{\top} D_me^{Q_m u} e_{i^{\prime}} du.
\end{eqnarray*}

 The second sum in (\ref{eq:sum1}) can be worked out as follows.
\begin{eqnarray*}
\mathbb{E}_{\theta}\Big[ \sum_{\ell=1}^{n-1} \sum_{r=0}^{\ell -1} \Phi_{k,m}X_{\ell}^k(i,j) X_{r}^k(i^{\prime},j^{\prime})\Big] &=&  \sum_{\ell=1}^{n-1} \sum_{r=0}^{\ell -1} \mathbb{E}_{\theta}\Big[
 \Phi_{k,m}X_{\ell}^k(i,j) X_{r}^k(i^{\prime},j^{\prime})\Big]\\
 &&\hspace{-5.5cm}=  \sum_{\ell=1}^{n-1} \sum_{r=0}^{\ell -1} 
 \mathbb{P}_{\theta}\big\{X_{\ell h}^k=i, X_{(\ell+1)h}^k=j, X_{rh}^k= i^{\prime}, X_{(r+1)h}^k=j^{\prime}, \Phi_{k,m}=1\big\}\\
  &&\hspace{-5.5cm}=  \sum_{\ell=1}^{n-1} \sum_{r=0}^{\ell -1} \sum_{v\in S}
 \mathbb{P}_{\theta}\big\{X_{\ell h}^k=i, X_{(\ell+1)h}^k=j, X_{rh}^k= i^{\prime}, X_{(r+1)h}^k=j^{\prime}, \Phi_{k,m}=1,X_0^k=v\big\}\\
  &&\hspace{-5.5cm}=  \sum_{\ell=1}^{n-1} \sum_{r=0}^{\ell -1} \sum_{v\in S} \mathbb{P}_{\theta}\{X_0^k=v\}\mathbb{P}_{\theta}\{\Phi_{k,m}=1\big\vert X_0^k=v\}\mathbb{P}_{\theta}\{X_{rh}^k=i^{\prime}\big\vert \Phi_{k,m}=1, X_0^k=v\}\\
  &&\hspace{-3.5cm}\times \mathbb{P}_{\theta}\{X_{(r+1)h}^k=j^{\prime}\big\vert \Phi_{k,m}=1, X_{rh}^k=i^{\prime}, X_0^k=v\}\\
   &&\hspace{-3.5cm}\times \mathbb{P}_{\theta}\{X_{\ell h}^k=i\big\vert \Phi_{k,m}=1, X_{(r+1)h}^k=j^{\prime}, X_{rh}^k = i^{\prime}, X_0^k=v\}\\
     &&\hspace{-3.5cm}\times  \mathbb{P}_{\theta}\{X_{(\ell +1) h}^k=j\big\vert \Phi_{k,m}=1, X_{\ell h}^k=i, X_{(r+1)h}^k=j^{\prime}, X_{rh}^k = i^{\prime}, X_0^k=v\}\\
  &&\hspace{-5.5cm}=  \sum_{\ell=1}^{n-1} \sum_{r=0}^{\ell -1} \sum_{v\in S} \pi_v \phi_{v,m}e_v^{\top} e^{Q_m rh} e_{i^{\prime}}  h e_{j^{\prime}}^{\top} e^{Q_m( \ell h -(r+1)h)} e_i  
  q_{ij,m}h\\
      &&\hspace{-5.5cm}=   q_{ij,m}q_{i^{\prime}j^{\prime},m} \sum_{\ell=1}^{n-1} \sum_{r=0}^{\ell -1} \pi^{\top}D_m  e^{Q_m rh} e_{i^{\prime}}  h e_{j^{\prime}}^{\top} e^{Q_m( \ell h -(r+1)h)} e_i  h
\end{eqnarray*}
which by dominated convergence theorem the last sum converges: 
\begin{eqnarray*}
&& q_{ij,m}q_{i^{\prime}j^{\prime},m} \sum_{\ell=1}^{n-1} \sum_{r=0}^{\ell -1} \pi^{\top}D_m  e^{Q_m rh} e_{i^{\prime}}  h e_{j^{\prime}}^{\top} e^{Q_m( \ell h -(r+1)h)} e_i  h \\
 &&\hspace{2cm}   \stackrel{h \rightarrow 0}{\longrightarrow} q_{ij,m}q_{i^{\prime}j^{\prime},m}   \int_0^T \int_0^u \pi^{\top}D_me^{Q_m t} e_{i^{\prime}} e_{j^{\prime}}^{\top} e^{Q_m (u-t)} e_i dt du.
 \end{eqnarray*}
Similarly, following the same arguments, one can show that the third sum in (\ref{eq:sum1})
\begin{eqnarray*}
&&\mathbb{E}_{\theta}\Big[ \sum_{r=1}^{n-1} \sum_{\ell=0}^{r -1} \Phi_{k,m}X_{\ell}^k(i,j) X_{r}^k(i^{\prime},j^{\prime})\Big]\\
&&\hspace{2cm}  \stackrel{h \rightarrow 0}{\longrightarrow}
 q_{ij,m}q_{i^{\prime}j^{\prime},m}   \int_0^T \int_0^u  \pi^{\top}D_m e^{Q_m t} e_{i} e_{j}^{\top} e^{Q_m (u-t)} e_{i^{\prime}} dt du.
\end{eqnarray*}
Putting the three limiting integrals together yields the third statement. $\blacksquare$

\subsubsection*{Proof of (\ref{eq:1d})}

Applying Fubini's theorem we get
\begin{eqnarray*}
\mathbb{E}_{\theta}\big[\Phi_{k,m} T_i^k T_{i^{\prime}}^k\big]&=& \int_0^T \int_0^T \mathbb{P}_{\theta}\big\{X_t^k=i, X_u^k=i^{\prime}, \Phi_{k,m}=1\big\}dtdu\\
&=&  \int_0^T \int_0^u \mathbb{P}_{\theta}\big\{X_t^k=i, X_u^k=i^{\prime}, \Phi_{k,m}=1\big\}dtdu\\
&&\hspace{1cm}+ \int_0^T \int_u^T \mathbb{P}_{\theta}\big\{X_t^k=i, X_u^k=i^{\prime}, \Phi_{k,m}=1\big\}dtdu.
\end{eqnarray*}
The first double integral can be worked out as follows
\begin{eqnarray*}
 &&\int_0^T \int_0^u \mathbb{P}_{\theta}\big\{X_t^k=i, X_u^k=i^{\prime}, \Phi_{k,m}=1\big\}dtdu\\
 &&\hspace{1cm}=
  \int_0^T \int_0^u \sum_{v\in S} \mathbb{P}_{\theta}\big\{X_t^k=i, X_u^k=i^{\prime}, \Phi_{k,m}=1, X_0^k=v\big\}dtdu\\
   &&\hspace{1cm}=
  \int_0^T \int_0^u \sum_{v\in S} \mathbb{P}_{\theta}\{X_0^k=v\}\mathbb{P}_{\theta}\{\Phi_{k,m}=1\big\vert X_0^k=v\}\mathbb{P}_{\theta}\big\{X_t^k=i\big\vert \Phi_{k,m}=1, X_0^k=v\big\}\\
  &&\hspace{2cm}\times \mathbb{P}_{\theta}\{ X_u^k=i^{\prime}\big\vert \Phi_{k,m}=1, X_t^k=i, X_0^k=v\}dtdu\\
   &&\hspace{1cm}=  \int_0^T \int_0^u \sum_{v\in S} \pi_v \phi_{v,m} e_v^{\top} e^{Q_m t} e_i e_i^{\top} e^{Q_m (u-t)} e_{i^{\prime}} dt du\\
     &&\hspace{1cm}=  \int_0^T \int_0^u \pi^{\top}D_m e^{Q_m t} e_i e_i^{\top} e^{Q_m (u-t)} e_{i^{\prime}} dt du.
\end{eqnarray*}
By the same lines of arguments, we get 
\begin{eqnarray*}
&&\int_0^T \int_u^T \mathbb{P}_{\theta}\big\{X_t^k=i, X_u^k=i^{\prime}, \Phi_{k,m}=1\big\}dtdu\\
 &&\hspace{2cm}=  \int_0^T \int_u^T \pi^{\top}D_m e^{Q_m t} e_{i^{\prime}} e_{i^{\prime}}^{\top} e^{Q_m (u-t)} e_{i} dt du\\
  &&\hspace{2cm}=  \int_0^T \int_0^t \pi^{\top}D_m e^{Q_m u} e_{i^{\prime}} e_{i^{\prime}}^{\top} e^{Q_m (t-u)} e_{i} dudt,
\end{eqnarray*}
where the last integral was due to using change of variable. Thus, the claim on the fourth statement is established. 
$\blacksquare$

\subsubsection*{Proof of (\ref{eq:1e})}

For notational convenience, define for $j\in S$, $u>0$, $Y_j^k(u)=\mathbf{1}_{\{X_u^k=j\}}$.
\begin{eqnarray}
&&\mathbb{E}_{\theta}\big[\Phi_{k,m} N_{ij}^k T_{i^{\prime}}^k\big]=\mathbb{E}_{\theta}\Big[\sum_{\ell=0}^{n-1} \int_0^T \Phi_{k,m} Y_{i^{\prime}}^k(u) X_{\ell}^k(i,j) du\Big]  \nonumber\\
&&\hspace{0cm}=\sum_{\ell =0}^{n-1} \int_0^T \mathbb{P}_{\theta}\{X_{\ell h}^k=i, X_{(\ell +1)h}^k=j, X_u^k=i^{\prime}, \Phi_{k,m}=1\}du \nonumber\\
&&\hspace{0cm}=\sum_{\ell =0}^{n-1} \int_0^T \sum_{v\in S}  \mathbb{P}_{\theta}\{X_{\ell h}^k=i, X_{(\ell +1)h}^k=j, X_u^k=i^{\prime}, \Phi_{k,m}=1, X_0^k=v\}du  \nonumber\\
&&\hspace{0cm}=\sum_{\ell =0}^{n-1} \int_0^{\ell h} \sum_{v\in S}  \mathbb{P}_{\theta}\{X_{\ell h}^k=i, X_{(\ell +1)h}^k=j, X_u^k=i^{\prime}, \Phi_{k,m}=1, X_0^k=v\}du  \nonumber\\
&&\hspace{0.5cm}+ \sum_{\ell =0}^{n-1} \int_{\ell h}^{(\ell+1)h} \sum_{v\in S}  \mathbb{P}_{\theta}\{X_{\ell h}^k=i, X_{(\ell +1)h}^k=j, X_u^k=i^{\prime}, \Phi_{k,m}=1, X_0^k=v\}du \nonumber\\
&&\hspace{0.5cm}+ \sum_{\ell =0}^{n-1} \int_{(\ell+1)h}^{T} \sum_{v\in S}  \mathbb{P}_{\theta}\{X_{\ell h}^k=i, X_{(\ell +1)h}^k=j, X_u^k=i^{\prime}, \Phi_{k,m}=1, X_0^k=v\}du. \nonumber
\end{eqnarray}
We evaluate each term in the last equality one by one. For the first term,
\begin{eqnarray*}
&&\sum_{\ell =0}^{n-1} \int_0^{\ell h} \sum_{v\in S}  \mathbb{P}_{\theta}\{X_{\ell h}^k=i, X_{(\ell +1)h}^k=j, X_u^k=i^{\prime}, \Phi_{k,m}=1, X_0^k=v\}du\\
&&\hspace{0cm}= \sum_{\ell =0}^{n-1} \int_0^{\ell h} \sum_{v\in S}  \mathbb{P}_{\theta}\{X_0^k=v\}\mathbb{P}_{\theta}\{\Phi_{k,m}=1\big\vert X_0^k=v\}\mathbb{P}_{\theta}\{X_u^k=i^{\prime}\big\vert \Phi_{k,m}=1, X_0^k=v\}\\
&&\hspace{2cm}\times
\mathbb{P}_{\theta}\{X_{\ell h}^k=i\big\vert \Phi_{k,m}=1, X_u^k=i^{\prime}, X_0^k=v\} \\
&&\hspace{2cm}\times
\mathbb{P}_{\theta}\{X_{(\ell+1)h}^k=j\big\vert \Phi_{k,m}=1,X_{\ell h}^k=i, X_u^k=i^{\prime}, X_0^k=v\}\\
&&= \sum_{\ell =0}^{n-1} \int_0^{\ell h} \sum_{v\in S} \pi_v \phi_{v,m} e_v^{\top} e^{Q_m u} e_{i^{\prime}} e_{i^{\prime}}^{\top} e^{Q_m (\ell h -u)} e_i q_{ij,m} h du\\
&&= \sum_{\ell =0}^{n-1} \int_0^{\ell h} \pi^{\top} D_m e^{Q_m u} e_{i^{\prime}} e_{i^{\prime}}^{\top} e^{Q_m (\ell h -u)} e_i q_{ij,m} h du\\
&& \stackrel{h\rightarrow 0}{\longrightarrow} q_{ij,m}\int_0^T \int_0^t   \pi^{\top} D_m e^{Q_m u} 
 e_{i^{\prime}} e_{i^{\prime}}^{\top} e^{Q_m(t-u)} e_i du dt,
\end{eqnarray*}
where the convergence is due to dominated convergence theorem. Furthermore,
\begin{eqnarray*}
 &&\sum_{\ell =0}^{n-1} \int_{\ell h}^{(\ell+1)h} \sum_{v\in S}  \mathbb{P}_{\theta}\{X_{\ell h}^k=i, X_{(\ell +1)h}^k=j, X_u^k=i^{\prime}, \Phi_{k,m}=1, X_0^k=v\}du\\
&&\hspace{1cm}= \sum_{\ell =0}^{n-1} \int_{\ell h}^{(\ell+1)h} \sum_{v\in S} \mathbb{P}_{\theta}\{X_0^k=v\}\mathbb{P}_{\theta}\{\Phi_{k,m}=1\big\vert X_0^k=v\}\\
&&\hspace{2cm}\times \mathbb{P}_{\theta}\{X_{\ell h}=i\big\vert \Phi_{k,m}=1,X_0^k=i\}\\
&&\hspace{2cm}\times \mathbb{P}_{\theta}\{X_u^k=i^{\prime}\big\vert \Phi_{k,m}=1, X_{\ell h}^k=i,X_0^k=v\}\\
&&\hspace{2cm}\times \mathbb{P}_{\theta}\{X_{(\ell+1)h}^k=j\big\vert \Phi_{k,m}=1,X_u^k=i^{\prime},X_{\ell h}^k=i,X_0^k=v\}du\\
&&\hspace{1cm}= \sum_{\ell =0}^{n-1} \int_{\ell h}^{(\ell+1)h} \sum_{v\in S} \pi_v \phi_{v,m} e_i^{\top} e^{Q_m \ell h} e_i e_i^{\top} e^{Q_m(u-\ell h)} e_{i^{\prime}} e_{i^{\prime}}^{\top} e^{Q_m((\ell+1)h- u)} e_j du\\
&&\hspace{1cm}= \sum_{\ell =0}^{n-1} \int_{\ell h}^{(\ell+1)h} \pi^{\top} D_m e^{Q_m \ell h} e_i e_i^{\top} e^{Q_m(u-\ell h)} e_{i^{\prime}} e_{i^{\prime}}^{\top} e^{Q_m((\ell+1)h- u)} e_j du\\
&&\hspace{1cm}=\mathcal{O}(h).
\end{eqnarray*}
The last equality is due to fact that for $\ell h\leq u <(\ell+1)h$, $e_i^{\top} e^{Q_m (u-\ell h)} e_{i^{\prime}}\approx e_i^{\top}\big[ I + Q_m h\big]e_{i^{\prime}}=q_{i i^{\prime},m}h $ for small $h>0$. Similarly, $e_{i^{\prime}}^{\top} e^{Q_m ((\ell+1)h-u)} e_{j}\approx q_{i^{\prime} j,m}h$. Thus, the contribution of the second term in the decomposition of $\mathbb{E}_{\theta}[\Phi_{k,m}N_{ij}^k T_{i^{\prime}}^k]$ is negligible. Finally, the third term is evaluated as follows.
\begin{eqnarray*}
&&\sum_{\ell =0}^{n-1} \int_{(\ell+1)h}^{T} \sum_{v\in S}  \mathbb{P}_{\theta}\{X_{\ell h}^k=i, X_{(\ell +1)h}^k=j, X_u^k=i^{\prime}, \Phi_{k,m}=1, X_0^k=v\}du\\
&&\hspace{1cm}= \sum_{\ell =0}^{n-1} \int_{(\ell+1)h}^{T} \sum_{v\in S} \mathbb{P}_{\theta}\{X_0^k=v\}\mathbb{P}_{\theta}\{\Phi_{k,m}=1\big\vert X_0^k=v\}\\
&&\hspace{2.5cm}\times \mathbb{P}_{\theta}\{X_{\ell h}^k = i\big\vert \Phi_{k,m}=1,X_0^k=v\}\\
&&\hspace{2.5cm}\times \mathbb{P}_{\theta}\{ X_{(\ell+1)h}^k=j\big\vert \Phi_{k,m}=1,X_{\ell h}^k=i, X_0^k=v\}\\
&&\hspace{2.5cm}\times \mathbb{P}_{\theta}\{X_u^k=i^{\prime}\big\vert \Phi_{k,m}=1,X_{(\ell+1)h}^k=j, X_{\ell h}^k=i, X_0^k=v\}\\
&&\hspace{1cm}= \sum_{\ell =0}^{n-1} \int_{(\ell+1)h}^{T} \sum_{v\in S} \pi_v \phi_{v,m} e_v^{\top} e^{Q_m \ell h} e_i q_{ij,m}h e_j^{\top} e^{Q_m(u-(\ell+1)h)} e_{i^{\prime}} du\\
&&\hspace{1cm}= q_{ij,m}\sum_{\ell =0}^{n-1} \int_{(\ell+1)h}^{T} \pi^{\top} D_m e^{Q_m \ell h} e_i  e_j^{\top} e^{Q_m(u-(\ell+1)h)} e_{i^{\prime}} h du\\
&&\hspace{1cm}\stackrel{h\rightarrow 0}{\longrightarrow}
q_{ij,m}\int_0^T \int_t^T \pi^{\top} D_m e^{Q_m t} e_i  e_j^{\top} e^{Q_m(u-t)} e_{i^{\prime}} dudt\\
&&\hspace{1cm}=q_{ij,m}\int_0^T \int_0^u \pi^{\top} D_m e^{Q_m t} e_i  e_j^{\top} e^{Q_m(u-t)} e_{i^{\prime}} dtdu,
\end{eqnarray*}
which in turn establishes the claim for the fifth statement. $\blacksquare$

\subsection{Derivation of the covariance matrix $\Sigma(\theta^0)$}


If the EM estimator $\widehat{\theta }$ converges to the true value $\theta $, it follows from Chernoff (1956) and Cramer (1946) p. 254 that for each $%
(i,j)\in S$ and $1\leq m\leq M$ the random variable $\widehat{N}_{ij,m}/K-N_{ij,m}/K$, which is equal to $\frac{1}{K}%
\sum_{k=1}^{K}\widehat{\Phi }_{k,m}N_{ij}^{k}-\frac{1}{K}\sum_{k=1}^{K}\Phi
_{k,m}N_{ij}^{k}$, converges to zero in probability as $K\rightarrow \infty $. It follows, by independence of $\{X^{k}\}$, and above references, $\frac{1}{K}\sum_{k=1}^{K}\mathbb{E}_{\theta}\big[\Phi _{k,m}\big\vert X^{k}\big]N_{ij}^{k}=\frac{1}{K}\sum_{k=1}^{K}\mathbb{E}_{\theta}\big[\Phi
_{k,m}N_{ij}^{k}|X^{k}\big]\stackrel{\mathbb{P}_{\theta}}{\Longrightarrow }\mathbb{E}_{\theta}\big[\Phi _{k,m}N_{ij}^{k}\big]$ which is the same convergence as $\frac{1}{K%
}\sum_{k=1}^{K}\Phi _{k,m}N_{ij}^{k}$. This implies that $\widehat{N}%
_{ij,m}/K$ and $N_{ij,m}/K$ have the same distributional convergence. See
van der Vaart (2000), Ch. 2 on stochastic convergence of random variables.
The same reasoning applies to $\widehat{T}_{i,m}/K-T_{i,m}/K$ and 
$\widehat{B}_{i,m}/K-B_{i,m}/K$. As $\widehat{\theta}\rightarrow \theta$, it follows from the statement above that
\begin{eqnarray*}
\sqrt{K}\big(\widehat{q}_{ij,m}-q_{ij,m}\big)=\frac{\sqrt{K}}{\widehat{T}_{i,m}/K}\Big(\widehat{N}_{ij,m}/K-q_{ij,m}\widehat{T}_{i,m}/K\Big)
\end{eqnarray*}
has the same asymptotic distribution as the random variable
\begin{eqnarray*}
\frac{\sqrt{K}}{\mathbb{E}_{\theta}\big[\Phi_{k,m}T_i^k\big]}\Big(N_{ij,m}/K-q_{ij,m}T_{i,m}/K\Big).
\end{eqnarray*}
By the multivariate central limit theorem, the above has asymptotic multivariate normal distribution with mean zero and the covariance matrix
\begin{eqnarray*}
\frac{\mathbb{E}_{\theta}\Big[\big(\Phi_{k,n}N_{i^{\prime}j^{\prime}}^k -q_{i^{\prime}j^{\prime},n}\Phi_{k,n}T_{i^{\prime}}^k\big)\big(\Phi_{k,m}N_{ij}^k -q_{ij,m}\Phi_{k,m}T_i^k \big)\Big]}{\mathbb{E}_{\theta}[\Phi_{k,n}T_{i^{\prime}}^k]\mathbb{E}_{\theta}[\Phi_{k,m}T_i^k]}.
\end{eqnarray*}
Notice that we have used the fact that $\mathbb{E}_{\theta}\Big[\Phi_{k,n}N_{i^{\prime}j^{\prime}}^k -q_{ij,m}\Phi_{k,n}T_{i^{\prime}}^k\Big]=0$, see Lemma \ref{lem:lemB2}. By Proposition \ref{prop:B1} we have
\begin{eqnarray*}
&&\mathbb{E}_{\theta}\Big[\big(\Phi_{k,n}N_{i^{\prime}j^{\prime}}^k -q_{i^{\prime}j^{\prime},n}\Phi_{k,n}T_{i^{\prime}}^k\big)\big(\Phi_{k,m}N_{ij}^k -q_{ij,m}\Phi_{k,m}T_i^k \big)\Big]\\
&&\hspace{1.5cm}=\delta_m(n)\mathbb{E}_{\theta}\big[\Phi_{k,m}N_{ij}^k N_{i^{\prime}j^{\prime}}^k\big]-\delta_m(n)q_{i^{\prime}j^{\prime},n}\mathbb{E}_{\theta}\big[\Phi_{k,n}N_{ij}^kT_{i^{\prime}}^k\big]\\
&&\hspace{2cm}+\delta_m(n)q_{ij,m}\mathbb{E}_{\theta}\big[\Phi_{k,n}N_{i^{\prime}j^{\prime}}^k T_i^k\big] + q_{ij,m}q_{i^{\prime}j^{\prime},n}\delta_m(n)\mathbb{E}_{\theta}\big[\Phi_{k,n}T_{i^{\prime}}^k T_i^k\big]\\
&&\hspace{1.5cm}=\delta_m(n)\delta_i(i^{\prime})\delta_j(j^{\prime})q_{ij,n}\int_0^T \pi^{\top}D_n e^{Q_n u} e_i du,
\end{eqnarray*}
which establishes the assertion about $\textrm{Cov}\big(\widehat{q}_{i^{\prime}j^{\prime},n}-q_{i^{\prime}j^{\prime},n}, \widehat{q}_{ij,m}-q_{ij,m}\big)$ noting that
\begin{eqnarray*}
\mathbb{E}_{\theta}\big[\Phi_{k,n}T_i^k\big]=\int_0^T \pi^{\top}D_n e^{Q_n u} e_i du.
\end{eqnarray*}

By similar arguments, one can show subsequently using Proposition \ref{prop:B1},
\begin{eqnarray*}
\textrm{Cov}\big(\widehat{\phi}_{i^{\prime},n}-\phi_{i^{\prime},n},\widehat{q}_{ij,m}-q_{ij,m}\big)
=
\frac{\mathbb{E}_{\theta}\Big[\big(\Phi_{k,n}B_{i^{\prime}}^k-\phi_{i^{\prime},n}B_{i^{\prime}}^k\big)\big(\Phi_{k,m}N_{ij}^k -q_{ij,m}\Phi_{k,m}T_i^k\big)\Big]}{\mathbb{E}_{\theta}\big[B_{i^{\prime}}^k\big]\mathbb{E}_{\theta}\big[\Phi_{k,m}T_i^k\big]}
=0,
\end{eqnarray*}
and
\begin{eqnarray*}
\textrm{Cov}\big(\widehat{\phi}_{i^{\prime},n}-\phi_{i^{\prime},n},\widehat{\phi}_{i,m}-\phi_{i,m}\big)&=&
\frac{\mathbb{E}_{\theta}\Big[\big(\Phi_{k,n}B_{i^{\prime}}^k-\phi_{i^{\prime},n}B_{i^{\prime}}^k\big)(\big(\Phi_{k,m}B_{i}^k-\phi_{i,m}B_{i}^k\big)\Big]}{\mathbb{E}_{\theta}\big[B_{i^{\prime}}^k\big]\mathbb{E}_{\theta}\big[B_{i}^k\big]}\\
&=&\frac{\phi_{i^{\prime},n}}{\pi_i^{\prime}}\delta_i(i^{\prime})\big(\delta_m(n)-\phi_{i,m}\big),
\end{eqnarray*}
which complete the proof of the theorem. $\blacksquare$

\medskip

\noindent The proof for absorption time $T$ follows very similar arguments to those in the proof for fixed $T$.

\section{Proof of Proposition 4}

Since the observed Fisher information matrix $I(\theta)$ presented in (17) of the paper, simplifies the conditional expectation of outer product of the complete-data score function, the elements $I(\theta_i,\theta_j)$ of the matrix $I(\theta)$ can be derived in straightforward way.

\subsubsection*{Elements of the intensity matrix $I(\phi _{j,n},\phi _{i,m})$}

To obtain the elements $I(\phi _{j,n},\phi _{i,m})$ of $%
I(\widehat{\phi }),$ we compute the first and second order derivatives of $%
\log L_{c}^k(\phi )$ for $(i,j)\in E$ and $1\leq m,n\leq M-1$: 
\begin{eqnarray}
\mathbb{E}_{\theta }\left[ -\frac{\partial ^{2}\log L_{c}^k(\phi )}{\partial
\phi _{j,n}\partial \phi _{i,m}}\Big\vert X^k\right]  &=&
\mathbb{E}_{\theta }\left[\left(\frac{\Phi_{k,n}}{\phi_{j,n}^2}\delta_m(n)+\frac{\Phi_{k,M}}{\phi_{j,M}^2}\right)B_j^k\delta_i(j)\Big\vert X^k\right] \nonumber \\
&=& \left(\frac{\widehat{\Phi}_{k,n} B_j^k}{\phi_{j,n}^2}\delta_m(n)+\frac{\widehat{\Phi}_{k,M} B_j^k}{\phi_{j,M}^2}\right)\delta_i(j). \nonumber 
\end{eqnarray}%
Using the score function with respect to $\phi_{i,m}$, see (7) of the paper, and since $B_i^kB_j^k=B_j^k\delta_i(j)$, $\Phi_{k,m}\Phi_{k,n}=\Phi_{k,n}\delta_m(n)$, and $\delta_M(m)=0$ for $m\neq M$,
\begin{eqnarray}
&&\mathbb{E}_{\theta }\left[ \left(\frac{\partial \log L_{c}^k(\phi )}{\partial
\phi _{j,n}}\right)\left(\frac{\partial \log L_{c}^k(\phi )}{\partial \phi _{i,m}}%
\right)\Big\vert X^k\right]   \nonumber \\
&&\hspace{2.5cm}=\mathbb{E}_{\theta }\left[\left(\frac{\Phi_{k,m}B_i^k}{\phi_{i,m}}-\frac{\Phi_{k,M}B_i^k}{\phi_{i,M}}\right)\left(\frac{\Phi_{k,n}B_j^k}{\phi_{j,n}}-\frac{\Phi_{k,M}B_j^k}{\phi_{j,M}}\right)\Big\vert X^k\right]  \nonumber \\[5pt]
&&\hspace{2.5cm}=\mathbb{E}_{\theta }\left[\frac{\Phi_{k,n}B_j^k}{\phi_{i,m}\phi_{j,n}}\delta_i(j)\delta_m(n) + \frac{\Phi_{k,M}B_j^k}{\phi_{i,M}\phi_{j,M}}\delta_i(j) \Big\vert X^k\right] \nonumber\\
&&\hspace{2.5cm}=\frac{\widehat{\Phi}_{k,n}B_j^k}{\phi_{i,m} \phi_{j,n}}\delta_i(j)\delta_m(n) + \frac{\widehat{\Phi}_{k,M}B_j^k}{\phi_{i,M}\phi_{j,M}}\delta_i(j). \nonumber 
\end{eqnarray}%
Following the score function for $\phi_{i,m}$, for $i\in E$ and $1\leq m\leq M-1$, we have
\begin{eqnarray}
\mathbb{E}_{\theta }\left[ \frac{\partial \log L_{c}^k(\phi )}{\partial
\phi _{i,m}}\Big\vert X^k\right] &=&\mathbb{E}_{\theta}\left[\frac{\Phi_{k,m}B_i^k}{\phi_{i,m}}-\frac{\Phi_{k,M}B_i^k}{\phi_{i,M}}\Big\vert X^k\right] \nonumber\\
&=& \frac{\widehat{\Phi}_{k,m}B_i^k}{\phi_{i,m}}-\frac{\widehat{\Phi}_{k,M}B_i^k}{\phi_{i,M}}. \label{5b2}
\end{eqnarray}
Replacing the parameters $\phi_{i,m}$ and $\phi_{j,n}$ by their respective estimates $\widehat{\phi}_{i,m}$ and $\widehat{\phi}_{j,n}$, the elements $I(\widehat{\phi}_{i,m},\widehat{\phi}_{j,n})$ is obtained by taking the sum $\sum_{k=1}^K$ of all the three pieces, given by
\begin{eqnarray*}
I(\widehat{\phi}_{i,m},\widehat{\phi} _{j,n})=\frac{\delta _{i}(j)}{\widehat{\phi }_{j,n}%
\widehat{\phi }_{i,m}}\sum_{k=1}^{K}\widehat{\Psi }_{j,n\vert M}^k\widehat{\Psi }_{i,m\vert M}^kB_{j}^{k}. \;\;
\end{eqnarray*}
This completes the proof. $\blacksquare$ 

\subsubsection*{Elements of the intensity matrix $I(\widehat{q}_{rv,n},\widehat{q}_{ij,m})$} 

Since the first order partial derivative of the log-likelihood $\log L_c^k(q)$ with respect to the parameter $q_{ij,m}$ is given by $\frac{\partial \log L_c^k(q)}{\partial q_{ij,m}}=\frac{\Phi_{k,m}N_{ij}^k}{q_{ij,m}}-\Phi_{k,m}T_{i,m}$, we have
\begin{eqnarray}
\mathbb{E}_{\theta }\Big[-\frac{\partial^2 \log L_c^k(q)}{\partial q_{ij,m} \partial q_{rv,n}}\Big\vert X^k\Big]&=&\mathbb{E}_{\theta}\left[ \frac{\Phi_{k,n} N_{rv}^k}{q_{ij,m}q_{rv,n}} \delta_i(r)\delta_j(v)\delta_m(n)  \Big\vert X^k\right] \nonumber \\
&=&  \frac{ \widehat{\Phi}_{k,n} N_{rv}^k}{q_{ij,m}q_{rv,n}} \delta_i(r)\delta_j(v)\delta_m(n). \nonumber 
\end{eqnarray}
Furthermore, since $A_{rv,n}^k:=N_{rv,n}^k-q_{rv,n}T_{r,n}^k$ for $(r,v)\in S$, $1\leq n\leq M$, 
\begin{eqnarray}
\mathbb{E}_{\theta }\left[ \Big( \frac{\partial \log L_c^k(q)}{\partial q_{rv,n}}\Big) \Big( \frac{\partial \log L_c^k(q)}{\partial q_{ij,m}}\Big)    \Big\vert X^k\right] &=& \mathbb{E}_{\theta}\left[\Big(\frac{\Phi_{k,m} A_{ij,m}^k}{q_{ij,m}} \Big)  \Big(\frac{\Phi_{k,n} A_{rv,n}^k}{q_{rv,n}} \Big) \Big\vert X^k\right] \nonumber\\
&=&  \frac{\delta_m(n)}{q_{ij,m}q_{rv,n}} \widehat{\Phi}_{k,n} A_{ij,m}^k A_{rv,n}^k. \nonumber
\end{eqnarray}
Using the score function for $q_{ij,m}$, for $i\in E$, $j\in S$ and $1\leq m\leq M$, we obtain
\begin{eqnarray}\label{derv1b}
\mathbb{E}_{\theta}\Big[\frac{\partial \log L_c^k(q)}{\partial q_{ij,m}}\Big\vert X^k\Big]=\frac{\widehat{\Phi}_{k,m}A_{ij,m}^k}{q_{ij,m}}
\end{eqnarray}
from which the elements of information matrix $I(q_{rv,n},q_{ij,m})$ are obtained as
\begin{eqnarray*}
I(\widehat{q}_{ij,m},\widehat{q}_{rv,n})=\frac{\delta_i(r)\delta_j(v)\delta_m(n)}{\widehat{q}_{ij,m}\widehat{q}_{rv,n}}\widehat{N}_{rv,n} -\frac{1}{\widehat{q}_{ij,m}\widehat{q}_{rv,n}}
\sum_{k=1}^K \widehat{\Phi}_{k,n}(\delta_m(n)-\widehat{\Phi}_{k,m})\widehat{A}_{ij,m}^k \widehat{A}_{rv,n}^k,
\end{eqnarray*}
which completes the proof. $\blacksquare$

\subsubsection*{Elements of the intensity matrix $I(\widehat{\phi}_{i,m},\widehat{q}_{rv,n})$}

It is straightforward to check that $\frac{\partial \log L_c^k(\theta)}{\partial \phi_{i,m} \partial q_{rv,n}}=0$ which in turn yields
\begin{eqnarray*}
\mathbb{E}_{\theta }\left[\frac{\partial \log L_c(\theta)}{\partial \phi_{i,m} \partial q_{rv,n}}\Big\vert X^k\right]=0.
\end{eqnarray*}
Using the complete-information score functions of $\phi_{i,m}$ and $q_{rv,n}$, we have for $i \in E, (r,v)\in S$, $1\leq m\leq M-1$ and $1\leq n\leq M$,
\begin{eqnarray*}
&&\mathbb{E}_{\theta }\left[\Big(\frac{\partial \log L_c^k(\theta)}{\partial \phi_{i,m}}\Big)\Big(\frac{\partial \log L_c^k(\theta)}{\partial q_{rv,n}}\Big)\Big\vert X^k\right]\nonumber \\
&&\hspace{2.5cm}=\mathbb{E}_{\theta }\left[\Big(\frac{\Phi_{k,m}B_i^k}{\phi_{i,m}} -\frac{\Phi_{k,M}B_i^k}{\phi_{i,M}}\Big)\frac{\Phi_{k,n}A_{rv,n}^k}{q_{rv,n}}\Big\vert X^k\right] \\
&&\hspace{2.5cm}=\mathbb{E}_{\theta}\left[\frac{\delta_m(n)\Phi_{k,n}B_i^k A_{rv,n}^k}{\phi_{i,m}q_{rv,n}} - \frac{\delta_M(n)\Phi_{k,n}B_i^k A_{rv,n}^k}{\phi_{i,M}q_{rv,n}} \Big\vert X^k\right]\\
&&\hspace{2.5cm}=\frac{\delta_m(n)\widehat{\Phi}_{k,n}B_i^k A_{rv,n}^k}{\phi_{i,m}q_{rv,n}} -  \frac{\delta_M(n) \widehat{\Phi}_{k,n}B_i^k A_{rv,n}^k}{\phi_{i,M}q_{rv,n}}.
\end{eqnarray*}
Following (\ref{5b2}) and (\ref{derv1b}), we obtain
\begin{eqnarray*}
&&\mathbb{E}_{\theta }\left[ \frac{\partial \log L_{c}^k(\phi )}{\partial
\phi _{i,m}}\Big\vert X^k\right]\mathbb{E}_{\theta}\Big[\frac{\partial \log L_c^k(q)}{\partial q_{rv,n}}\Big\vert X^k\Big]\\
&&\hspace{2.5cm}=\Big( \frac{\widehat{\Phi}_{k,m}B_i^k}{\phi_{i,m}}-\frac{\widehat{\Phi}_{k,M}B_i^k}{\phi_{i,M}}\Big)\frac{\widehat{\Phi}_{k,n}A_{rv,n}^k}{q_{rv,n}}\\
&&\hspace{2.5cm}=\frac{\widehat{\Phi}_{k,m}\widehat{\Phi}_{k,n}B_i^k A_{rv,n}^k}{\phi_{i,m} q_{rv,n}}  - \frac{\widehat{\Phi}_{k,M}\widehat{\Phi}_{k,n}B_i^k A_{rv,n}^k}{\phi_{i,M} q_{rv,n}}.
\end{eqnarray*}
Adding the three pieces together and replacing $\phi_{i,m}$ and $q_{rv,n}$ by their respective ML estimates, we obtain 
\begin{eqnarray*}
I(\widehat{\phi}_{i,m},\widehat{q}_{rv,n})=-\sum_{k=1}^K \frac{\widehat{\Phi}_{k,n}B_i^k \widehat{A}_{rv,n}^k}{\widehat{\phi}_{i,m}\widehat{q}_{rv,n}}\Big[ \delta_m(n)-\frac{\widehat{\phi}_{i,m}}{\widehat{\phi}_{i,M}}\delta_M(n) + \frac{\widehat{\phi}_{i,m}}{\widehat{\phi}_{i,M}}\widehat{\Phi}_{k,M} -\widehat{\Phi}_{k,m} \Big],
\end{eqnarray*}
which ends the proof. $\blacksquare$

\section{Absorption probabilities for the two-regime g-mixture}

The absorption probabilities $f_{ij,m}$, $m=1,2,$ from state $i\in E=\{1,2\}$
into states $j\in \Delta =\{3,4\},$ are computed using the transition matrix
of an m'th discrete Markov chain embedded into an m'th\ Markov jump process,
see Theorem 11.6 in Ch.11 of Grinstead and Snell (1997). The transition
matrix of the m-th embedded Markov chain, denoted by $P_{m},$ is obtained
from the intensity matrix $Q_{m}$ of an m'th Markov jump process by 
\begin{equation*}
p_{ij,m}=\frac{q_{ij,m}}{|q_{ii,m}|},i\in E,j\in S,j\neq i,
\end{equation*}%
where $q_{ij,m}$ is an (i,j)th entry in the intensity matrix $Q_{m}.$ To
estimate $f_{ij,m},m=1,2,$ we use $\widehat{Q}_{m}$ from Section 6.3 to
first obtain $\widehat{P}_{m},$ the estimate of $P_{m}$: 
\begin{equation*}
\widehat{P}_{1}=%
\begin{array}{c}
1 \\ 
2%
\end{array}%
\left( 
\begin{array}{ccccc}
0 & 0.10283 & 0.89717 & 0.00000 &  \\ 
0.87269 & 0 & 0.10155 & 0.02576 & 
\end{array}%
\right) ,
\end{equation*}%
and%
\begin{equation*}
\widehat{P}_{2}=%
\begin{array}{c}
1 \\ 
2%
\end{array}%
\left( 
\begin{array}{ccccc}
0 & 0.12431 & 0.78503 & 0.09066 &  \\ 
0.50746 & 0 & 0.24046 & 0.25208 & 
\end{array}%
\right) ,
\end{equation*}%
where for notational convenience, the last two rows of zeros were deleted.
We then obtain the equations for the estimated absorption probabilities $%
\widehat{f}_{ij,m}$ in terms of $\widehat{P}_{m}$, $m=1,2$, see Exercise 34
in Ch.11 of Grinstead and Snell (1997): 
\begin{eqnarray*}
\widehat{f}_{13,m} &=&\widehat{p}_{13,m}+\widehat{p}_{12,m}\widehat{f}%
_{23,m}, \\
\widehat{f}_{14,m} &=&\widehat{p}_{14,m}+\widehat{p}_{12,m}\widehat{f}%
_{24,m},
\end{eqnarray*}%
and%
\begin{eqnarray*}
\widehat{f}_{23,m} &=&\widehat{p}_{23,m}+\widehat{p}_{21,m}\widehat{f}%
_{13,m}, \\
\widehat{f}_{24,m} &=&\widehat{p}_{24,m}+\widehat{p}_{21,m}\widehat{f}%
_{14,m},
\end{eqnarray*}%
from which we obtain $\widehat{f}_{ij,m},m=1,2,$ explicitly as 
\begin{eqnarray*}
\widehat{f}_{13,m} &=&\frac{\widehat{p}_{13,m}+\widehat{p}_{12,m}\widehat{p}%
_{23,m}}{1-\widehat{p}_{12,m}\widehat{p}_{21,m}}, \\
\widehat{f}_{14,m} &=&\frac{\widehat{p}_{14,m}+\widehat{p}_{12,m}\widehat{p}%
_{24,m}}{1-\widehat{p}_{12,m}\widehat{p}_{21,m}},
\end{eqnarray*}%
and 
\begin{eqnarray*}
\widehat{f}_{23,m} &=&\frac{\widehat{p}_{23,m}+\widehat{p}_{21,m}\widehat{p}%
_{13,m}}{1-\widehat{p}_{12,m}\widehat{p}_{21,m}}, \\
\widehat{f}_{24,m} &=&\frac{\widehat{p}_{24,m}+\widehat{p}_{21,m}\widehat{p}%
_{14,m}}{1-\widehat{p}_{12,m}\widehat{p}_{21,m}}.
\end{eqnarray*}

\nocite{*}
\bibliographystyle{acm}